\documentclass[11pt]{article}

\usepackage[margin=1in]{geometry}
\usepackage{setspace}
\usepackage[T1]{fontenc}
\usepackage[utf8]{inputenc}
\usepackage{lmodern}
\usepackage{microtype}
\usepackage{tikz}
\usetikzlibrary{patterns}
\usetikzlibrary{decorations.pathreplacing}

\usepackage{amsmath,amssymb,amsthm,mathtools,bm}

\usepackage{graphicx}
\usepackage{booktabs}
\usepackage{caption}
\usepackage{subcaption}
\usepackage{float}
\usepackage{rotating} 

\usepackage{enumitem}
\usepackage{xcolor}

\usepackage[round,authoryear]{natbib}
\usepackage{hyperref}
\hypersetup{
  colorlinks=true,
  linkcolor=blue,
  citecolor=blue,
  urlcolor=blue
}
\usepackage{cleveref} 

\numberwithin{equation}{section}
\newtheorem{theorem}{Theorem}[section]
\newtheorem{proposition}[theorem]{Proposition}
\newtheorem{lemma}[theorem]{Lemma}

\theoremstyle{definition}

\theoremstyle{remark}

\DeclareMathOperator{\E}{\mathbb{E}}

\title{Collateral and Reputation\\ in a Model of Strategic Defaults\thanks{I am indebted to my supervisor, Christian Hellwig, for guidance and encouragement throughout the project. I would also like to thank [full acknowledgments as in your file, including seminar participants and funding acknowledgement “Investissements d’Avenir” (LabEx Ecodec/ANR-11-LABX-0047)].}}

\author{Georgy Lukyanov\thanks{\'{E}cole Polytechnique, 5 av. Le Chatelier, 91120 Palaiseau, France. Email: georgy.lukyanov@polytechnique.edu}}

\date{}

\begin{document}
\maketitle



\medskip

\noindent
\begingroup
\setlength{\fboxsep}{8pt}\setlength{\fboxrule}{0.4pt}%
\fbox{%
  \begin{minipage}{0.97\textwidth}
  \small
  \textbf{Author's accepted manuscript (postprint)} of:\\
  Lukyanov, G. (2023). \emph{Collateral and reputation in a model of strategic defaults}. 
  \textit{Journal of Economic Dynamics \& Control}, 156, 104755. 
  \textbf{Version of Record (VoR) DOI:}
  \href{https://doi.org/10.1016/j.jedc.2023.104755}{10.1016/j.jedc.2023.104755}.\\[2pt]
  \textbf{License for this manuscript:} CC BY-NC-ND 4.0 
  \href{https://creativecommons.org/licenses/by-nc-nd/4.0/}{(link)}.
  \end{minipage}%
}
\endgroup

\medskip
%

\section{\label{intro}Introduction}

The length and severity of the U.S. 2007--2009 mortgage crisis alerted both the academic community and the wider public to its causes and to the damage widespread defaults inflicted on the economy as a whole. It was suggested that \emph{strategic defaults}\footnote{The findings of \cite{Fayetal2002} suggest that an important component affecting a household's decision to file for bankruptcy is the financial benefits of default. According to \cite{Edmans2010}, strategic defaults constitute around 30\% of all housing delinquencies.} constituted a significant fraction of all loan delinquencies.\footnote{See \cite{Zingales2010}.} Furthermore, the collapse in house prices and the associated reduction in home equity for the borrowers were a major reason for households to default on their mortgages. Even though ample evidence exists that a drop in home equity and a decision to default are tightly connected,\footnote{The good overview of the empirical evidence is provided by \cite{MianSufi2015}.} there remains a gap in the theoretical literature linking the borrower's reputation for repayment, on the one hand, and the use of collateral for lending, on the other. The present paper tries to fill in this gap.

We focus on the interaction between extrinsic and intrinsic incentives to repay. It is shown that there exists \emph{complementarity} between developing a reputation for being honest and pledging the asset as collateral: Other things being equal, the potential borrower would not wish to take a loan if either (i) he can develop a reputation, but cannot collateralize the loan, or (ii) can use collateral but cannot a build reputation. At the same time, the borrower would be willing to take a loan when \emph{both} channels are present.

This result is non-trivial, since common sense would suggest that reputation and collateral are \emph{substitutes} for each other: the borrower would either use his goodwill or physical collateral in the loan contract. Complementarity arises for the following reason: It is precisely when house prices fall that the borrower finds it most tempting to default on the loan -- and therefore, debt repayment signals to the lenders that the borrower is of the `honest' type, which in turn would allow borrowing on more favorable terms in the future. Hence, this reputational channel acts as \emph{insurance against downside risks} associated with asset price drops.

In addition, we explore how the option to sell the asset interacts with the option to default. The borrower would prefer to sell the asset whenever the ratio of its price relative to the borrower's non-financial income is either very low or very high, and the borrower would prefer to borrow against the asset if this ratio is intermediate.

This paper builds a setup in which the long-living borrower in each period applies for the loan from a group of short-lived lenders. Each lender observes whether the borrower has defaulted in the past and updates his posterior belief that the borrower is an `honest' type. The model demonstrates that asset ownership can significantly lever up the borrower's scope of reputation building: He is more eager to choose to borrow, as long as the ratio of the asset price to his non-financial income is neither too high nor too low. The opportunity to pledge the asset, in the form of physical collateral, offers additional credibility that is valuable, because it increases the possibility to develop his reputation: In the future, the asset price might drop, and if \emph{despite this drop}, the borrower still decides to repay, this acts as a \emph{signal} to the lenders that the borrower is `honest.' In that sense, physical and reputational collateral tend to be \emph{complementary} to each other: The asset becomes valuable both for the option to resell it in the future and for the option to default.

Reputation is introduced along the lines of \cite{KrepsWilson1982} and \cite{MilgromRoberts1982}. Each lender starts with a given prior probability that the borrower is `honest' (meaning that he faces infinitely high cost of default, and thus will repay the loan under all circumstances). That way, the `fully rational' borrower who \emph{is} capable of default (the `strategic' type) may find it worth \emph{mimicking} the behavior of the `honest' type via current repayment in order to benefit from default on a loan in the future.

The main body of the paper develops a three-period example in which the individual borrower faces overlapping generations of short-lived, perfectly competitive lenders. At each date, the borrower can offer a non-contingent debt contract. Each lender observes the proposed contract and the borrower's credit history; based on that information, the lender tries to infer the borrower's type and decides whether to accept the loan or reject it. Debt contracts can also be collateralized by the asset, whose price is assumed to fluctuate for exogenous reasons.

An important feature of our setup is that the borrower \emph{cannot commit} to the future debt contracts. At each date he engages in the short-term relationship with the current-period lenders, and we require his behavior to be sequentially rational. As we argue (in Section \ref{onlyreputation}), this constraint severely restricts the borrower's ability to build reputation. Pledging the asset as collateral can help the honest type overcome this commitment problem. The equilibrium structure developed in our setting points to the existence of \emph{complementarity} between the external and internal enforcement mechanisms for repayment: at some price levels, the borrower can gain more by establishing a reputation for honesty whenever he can also collateralize the loan contract. In this sense, reputational concerns complement physical collateral and vice versa.

The main results of the paper are stated in Proposition \ref{initdef} and Theorem \ref{mainresult}. They characterize equilibrium, in which at the initial date, the borrower prefers to sell the asset at very high and very low prices, but chooses to keep the asset and borrow against it for the intermediate price range. In the continuation game when there was borrowing at the initial date, the strategic borrower (i) defaults on the loan with certainty if the asset price turns out to be very low, (ii) repays the debt with certainty and sells the asset if the price turns out to be high, and (iii) randomizes between default and repayment whenever the price lies in between the two thresholds. Theorem \ref{mainresult} shows that the lower boundary for the borrower's initial reputation, above which he prefers to keep the asset and use it as collateral (rather than selling it), is lower than the corresponding boundary for the 'only reputation' case, provided that the ratio of the borrower's financial to non-financial income is neither too high nor too low. Furthermore, this range expands as the borrower becomes more impatient.

\subsection{Literature review}

Our paper delivers some insights which fit within the empirical literature of the 2007--2009 housing bubble. \cite{MianSufi2009} suggest that credit expansion of 2002 to 2005 was accompanied by a sharp \emph{decline} in relative income growth, in contrast to their positive comovement in the preceding decade. The analysis of the benchmark case developed  in section \ref{onlyreputation} suggests that the reputational channel does not work, and thus the complementarity between collateral and reputation starts to play a role, once income growth slows down: this is the time when the borrowers need the collateral in order to be willing to continue reputation building. By contrast, the reputational channel works on its own at times of high income growth.

\cite{Guisoetal2013} establish that one of the main contributing factors to the willingness to walk away from the mortgage is the relative equity shortfall (as a percentage of the value of the house): a one standard deviation increase in this measure is likely to raise the probability of strategic default by a quarter, this impact being reduced for those who lived in their house for more than 5 years. In our setup, the borrower would wish to back the loan by the asset when the reputation is initially low; on the other hand, honest borrowers with a high reputation can successfully separate themselves from the `strategic' types when the reputation is initially high. Although our model is cast in a setting with fixed time horizon, one of its implications is that conditional on no default, the borrower's reputation gradually increases over time, and therefore, those who live in their houses for a longer time, are more likely to be `honest', in the terminology of our paper.

Exploring the key drivers of mortgage defaults, \cite{Eluletal2010} have pointed to the borrower's \emph{illiquidity} as one of the main factors, along with negative equity. In our framework, the borrower's illiquidity can be proxied his non-financial income.\footnote{In their study, \cite{Eluletal2010} used credit card utilization rates as a proxy for illiquidity. Additionally, they show that county-level unemployment shocks also contribute to the default risk; the same applies for those having a second mortgage.} The result of section \ref{assetreputation} implies that lower non-financial income in the future is likely to narrow down the range for asset prices at which the strategic type will randomize between default and repayment, thereby reducing the scope for reputation building and increasing the threshold below which default will be certain.

Focusing on the demand-side explanations, \cite{DellAricciaetal2012} tested the relationship between borrowing constraints and credit expansion during the U.S. housing boom. They showed that the areas with faster credit growth were facing smaller loan denial rates, subsequently experiencing a spike in delinquency rates, the effect complementary to the supply-side channel of increased securitization and the rise in housing prices. In the context of the present paper, the main drivers for increased demand for borrowing are heavy discounting and the prospective rise in future income. The findings of Theorem \ref{mainresult} suggest that both the decline in the discount factor and the increase in the borrower's future income tend to expand the range of parameters (the lenders' prior $\pi_0$) for which the reputational channel works.

The results of \cite{GarmaiseNatividad2017} explore the relationship between worsening of credit rating and future access to credit. They found that for delinquent borrowers, the downgrade accounts for a 25\% to 65\% reduction in subsequent financing. Although our setup cannot properly account for the downgrades,\footnote{Due to the extremely conservative borrowing on behalf of the `honest' type (the one who is assumed to face infinite costs of default), non-repayment completely reveals the borrower's type, which in turn implies that there can be no \emph{partial} downgrade in equilibrium.} it lends some indirect support for these findings. The reputational channel creates an endogenous intertemporal link between the borrowing constraints: any factor that (i) facilitates future borrowing and (ii) makes it more desirable at the same time facilitates current lending, since the borrower's incentive to repay the loan is positively affected by the subsequent access to credit.

Some of the insights of our paper are applicable to secondary loan markets in general. \cite{Charietal2014} develop a dynamic adverse selection model showing that a small reduction in collateral values tends to generate sharp collapses in issuance of new secondary loans, with a similar impact on incentives to develop reputation. However, in their story, reputation is related to the quality of the borrowers' projects, which determines the \emph{ability} to repay the loan. The strategic defaults' story outlined in our work proposes a complementary explanation which drives the qualitatively similar dynamics.

Dynamic complementarity between secured and unsecured credit was considered by \cite{Ordonezetal2019}. They explore how borrowers tend to use collateral to raise a lot of debt in one period in order to pay it back in the next period to signal that their type is `high'. The characteristics of the available collateral, its pledgeability and volatility, constrain the leverage dynamics. They show that good borrowers can costlessly separate from the bad ones by deleveraging, but have to pay a higher interest rate when the uncertainty is higher. In contrast to their work, in the equilibrium that we construct, strategic borrowers try to pool with the honest types.

\indent
 
The paper is organized as follows. Section \ref{model} describes the basic framework. Section \ref{benchmarks} studies several benchmark cases. Section \ref{assetreputation} focuses on the interplay between extrinsic and intrinsic repayment incentives; it delivers the main result of the paper. Section \ref{robustness} analyses the implications of our model, provides several extensions and discusses robustness. Section \ref{conclusion} concludes.

All the proofs are relegated to the \hyperref[proofs]{Appendix}.

\section{\label{model}The model}

\subsection{Agents}

Consider an individual who lives for $(T+1)$ periods $0,1,2,\ldots,T$. His instantaneous utility of consumption is given by $u(c_t)$, with $u'>0$ and $u''\leq0$. Moreover, the individual discounts the future, so that his overall lifetime utility equals
\begin{equation}
U(\{c_t\}_{t=0}^T)=\sum_{t=0}^T\beta^tu(c_t),\quad\text{with }\beta<1.
\end{equation}

The individual receives deterministic and publicly observed income stream $\{y_t\}_{t=0}^T$. Since he is assumed to start with zero debt, and therefore without loss of generality, we take the date-0 income to be zero: $y_0=0$. In what follows, $y_t$ will be referred to as date-$t$ \emph{non-financial} income.

\subsection{Asset}

The individual starts with holding one unit of non-divisible physical asset which can be sold at a price $p_t$ at each date $t\in\{0,\ldots,T\}$.\footnote{In order to relate the price process $\{p_t\}$ to the asset's fundamental value, we can think of the asset as delivering a stochastic dividend $p_2$ at the end of date $T$. At each date $t$, there is an arrival of new information regarding this dividend that is reflected in the current asset price $p_t$.} The asset price follows the stochastic process satisfying the martingale property: for each $t=0,1,\ldots,T-1$,
\begin{equation}
p_{t+1}|p_t\sim F(\cdot|p_t),\quad\text{with $\int \widetilde{p}d  F(\widetilde{p}|p_t)$}\quad \text{and $p_0>0$ given.}
\end{equation}

\subsection{Lending contracts}

At each date $t$, the individual (henceforth ``the borrower'') faces the group of perfectly competitive lenders who have deep pockets and whose discount factor is normalized to one. Each date-$t$ lender lives for two consecutive periods: he gives out the loan at date $t$ and receives repayment at date $t+1$.

The borrower approaches one lender at random and offers him the contract $\mathcal{B}_t$. The date-$t$ contract specifies (i) the amount that the individual can borrow today, $b_t$; (ii) the repayment on the debt that is to be made to the lender on a subsequent date, $R_{t+1}$; and (iii) in case if the borrower still holds the asset, the loan contract specifies whether the borrower would like to pledge the asset as a collateral, $\kappa_t\in\{0,1\}$, where $\kappa_t=1$ stands for collateralized loan.

If the borrower pledges the asset as collateral, in case of default in period $t$ (on the loan taken at the preceding date), the lender expropriates the asset and resells it at a price $p_t$ to a third party.

Perfect competition on the part of lenders implies that the borrower has full bargaining power. Once the borrower has offered the contract $\mathcal{B}_t$, the lender decides whether to accept or reject the contract; in case of indifference, the lender can randomize between accepting and rejecting, with the acceptance probability denoted by $\alpha_t$.\footnote{This assumption is important for the construction of equilibrium in section \ref{assetreputation}. See the discussion that follows Proposition \ref{initdef}.} It is assumed that, if the borrower has been turned down, he cannot approach another lender; likewise, he cannot approach multiple lenders simultaneously.

Importantly, we assume that the borrower \emph{cannot commit} to the lending contract $\mathcal{B}_t$ offered to the lender at date $t$ before date $t$ actually arrives. In particular, the repayment $R_t$ on the date-$(t-1)$ cannot be made contingent on $b_t$, the loan that the borrower would be able to get at date $t$.

The date-$t$ \emph{public history} includes the borrower's past and prospective income stream $\{y_t\}_{t=0}^T$, the history of asset prices up to date $t$, $\{p_0,\ldots,p_t\}$, the amount of borrower's asset holdings $a_t\in\{0,1\}$ (that is, whether the asset has been sold or not), and the borrower's \emph{credit history} indicating whether the borrower has ever defaulted in the past. It is assumed that the lender's acceptance decision $\alpha$ can be made contingent on the date-$t$ public history and the contract $\mathcal{B}_t$ with which the borrower approaches.

\subsection{Reputation}

At the start of date 0, the lenders hold a prior $\pi_0\in[0,1]$ that the borrower is of the `honest', type. In terms of modelling, one can think of this type as the one who suffers an infinite disutility from default.\footnote{It is only in this sense in which he differs from the borrower whom we refer to as `strategic'. In particular, the `honest' borrower's behavior is not postulated ad hoc, but derived from his optimization, \emph{given} the additional restriction that default on the debt is not an option for him.} With complementary probability $1-\pi_0$, the lenders believe that the borrower's type is `strategic'. This implies that he does not face any costs of default, and hence will default on the debt whenever he finds it profitable to do so.

At each date $t=1,2,\ldots$, given the public history, the lenders start with the prior $\pi_{t-1}$ and update their beliefs to $\pi_t$ using Bayes' rule. Date-$t$ lenders can draw inference about the borrower's type upon observing the contract $\mathcal{B}_t$ that he proposes, which raises the possibility of signalling by contracts.

\subsection{Timing}

The timing within period $t$ is as follows. First, the borrower observes the current asset price $p_t$ and decides whether to repay the existing debt $R_t$ or default. Second, conditional on repayment, the borrower obtains the per-period dividend income $x$ and decides whether to sell or to keep the asset. Finally, the borrower approaches the lenders with the loan contract $\mathcal{B}_t=(b_t,R_{t+1},\kappa_t)$. The lender accepts (with probability $\alpha$) or rejects (with probability $1-\alpha$) the contract, and the borrower consumes the remaining amount.

The borrower's date-$t$ consumption is given by
\begin{equation}
\label{budgetconstraint}
c_t=y_t+(1-d_t)\big[(x+p_ts_t)a_t-R_t+\alpha b_t\big],
\end{equation}
where $d_t\in\{0,1\}$ is the default decision, $a_t\in\{0,1\}$ is the beginning-of-period asset holdings and $s_t\in\{0,1\}$, with $s_t\leq a_t$, is the selling decision.

The consumption is required to be non-negative: $c_t\geq0$.\footnote{As we will see, this condition imposes an implicit restriction on the amount that can be borrowed: since the `honest' type faces infinite costs when he is \emph{forced} into default, this type would prefer to borrow only to the extent that he can be certain to have non-negative consumption without defaulting under \emph{all possible circumstances}.}

\subsection{An example}

This section puts forth an example which will be used throughout the analysis in sections \ref{benchmarks} and \ref{assetreputation}.

First, we take $T=2$, so that there are three dates $t=0,1,2$.

Second, individual's utility is \emph{linear} in consumption:
\begin{equation*}
u(c_t)=c_t.
\end{equation*}

Third, the conditional distribution of date-$t$ asset price is uniform:
\begin{equation*}
\text{for $t=0,1$:}\quad p_{t+1}|p_t\sim\mathcal{U}[0,2p_t],\quad\text{with $p_0>0$ given.}
\end{equation*}

\section{\label{benchmarks}Benchmark cases}

In this section, we consider the two benchmark cases: (i) when the borrower may pledge the asset as collateral, but cannot build his reputation, since the lenders are certain that he is strategic, and (ii) when he can develop reputation for honesty, but at the same time does not have any physical asset to pledge as collateral. The main takeaway of this exercise would be to show how these two channels work separately.

First, we show that the best thing that the borrower can do when he only has the asset but has no reputation is to sell the asset right away (Proposition \ref{prop:borrow}).

Second, we show that the borrower would not wish to borrow against his reputation unless he already starts with a sufficiently high reputation (Proposition \ref{pardomcontr}).

Our main result, Theorem \ref{mainresult}, stated in section \ref{assetreputation}, claims that when the borrower is able to pledge the asset as collateral, this expands the set of initial reputation levels, $\pi_0$, above which he would prefer to keep the asset as collateral and borrow against it.

\subsection{\label{onlyasset}Borrowing against asset}

We start with the situation when the lenders know with certainty that the borrower is of the `strategic' type, so that $\pi_0=0$. Since loan repayment cannot be enforced by reputation, the lenders will accept only the collateralized loan contracts -- those with $\kappa_t=1$.\footnote{The only feasible loan contract without collateral is the `autarkic' contract $\mathcal{B}_t=(0,0,0)$.} Hence, once the agent sells the asset, he can no longer borrow.

The next proposition characterizes the individual's optimal selling, borrowing and default decisions.

\begin{proposition}
\label{prop:borrow}
At date $t=2$, if the individual has not defaulted and has not sold the asset before, he will default on the date-1 loan whenever the asset price falls below the threshold $\hat{p}_2=R_2$.

At date $t=1$, the individual who kept the asset and borrowed against it repays his loan and offers the contract
\begin{equation*}
(\hat{b}_1,\hat{R}_2)=(p_1,2p_1),
\end{equation*}
which is equivalent to selling the asset at a price $p_1$.

At date $t=0$, the individual offers the contract $(\hat{b}_0,\hat{R}_1)=(p_0,2p_0)$, which is equivalent to selling the asset at a price $p_0$.
\end{proposition}

\begin{proof}
See \ref{proof:borrow}.
\end{proof}

Observe that the equilibrium is characterized in the usual backward induction fashion, so that we specify the strategies for the continuation game when the agent has kept the asset until date 2 and when he has not defaulted at date 1. However, a close inspection of the optimal borrowing contract reveals that the promised repayment $\hat{R}_{t+1}$ equals the highest possible realization of the asset price $p_{t+1}$, which means that the best borrowing contract will trigger default at date $t+1$ with probability one. Such a strategy delivers exactly the same payoff as the one from selling the asset right away at a price $p_t$.

Therefore, we conclude that the asset will \emph{not} be valued for its role as a physical collateral unless the borrower can supplement it with his reputation.

\subsection{\label{onlyreputation}Borrowing against reputation}

In this section, we consider the case when the individual does not hold the asset, and thus any loan contract $\mathcal{B}_t$ that he may offer to the lenders has $\kappa_t=0$. At the same time, the lenders hold a positive prior belief $\pi_0>0$ that the borrower is \emph{honest}.

Since there is no asset, the lenders cannot recover anything if there was a default, and so in period $t$ they will be willing to lend positive amount \emph{only} to the borrower who has so far maintained his reputation. In turn, this implies not only that the borrower who defaulted before will not be granted a loan, but also that \emph{there cannot be signalling through contracts}.

The argument is simple: if there existed a pair of separating contracts (such that one of the contracts were picked by the honest type and the other by the strategic type), then the strategic type would reveal himself by picking the contract. However, in the absence of collateral, there does not exist any mutually profitable contract between the lenders and the borrower commonly known to be strategic (since he cannot secure the loan by committing to repay it). The only contract that is offered by the strategic type and that will be accepted by the lender is thus the \emph{autarkic} contract $(0,0)$. On the other hand, the mutually profitable contract between the lenders and the honest type \emph{does} exist. Given that the contract that is attractive for the honest type will be \emph{a fortiori} attractive for the strategic type (as he borrows on the same terms plus has the option to default), there is no way for the honest type to separate himself from the strategic type.

Hence, in each period $t$, provided that there was no default in the past, the two types of borrowers \emph{pool} on the contracts, with the strategic type mimicking the honest type: if in equilibrium in question the honest type ends up offering the autarkic contract $\mathcal{B}_t=(0,0,0)$, the strategic type can do no better than that, because any other contract would reveal his type, and hence will be rejected.

Since pledging the asset as collateral is not possible ($\kappa_t=0$), to economize on notation, we will refer to the date-$t$ typical loan contract as $(b_t,R_{t+1})$.

We consider the class of equilibria with the following structure:
\begin{enumerate}
\item At date $0$, the borrowers pool on the contract $(b_0,R_1)$;
\item At date $t=1$, the strategic borrower defaults on the loan $(b_0,R_1)$ with probability $\delta_1\in[0,1]$;
\item Conditional on repayment at date $1$, the borrowers pool on the contract $(b_1,R_2)$;
\item At date $t=2$, the strategic borrower defaults with probability one on $(b_1,R_2)$.
\end{enumerate}

The next lemma characterizes all possible equilibria depending on the parameters.

\begin{lemma}
\label{lemmasignal}
There exist several classes of signalling equilibria:
\begin{enumerate}
\item[I.] If $\pi_0<\beta^2$, the only equilibrium is
\begin{equation*}
b_t^{*}=R_{t+1}^{*}=0,\quad\text{for $t=0,1$},
\end{equation*}
which corresponds to autarky.
\item[II.] If $\beta^2\leq\pi_0<\beta$, then in addition to the autarkic equilibrium, there exists two other types of equilibria:
\begin{enumerate}
\item[(i)] Any sequence of contracts $\{(b_0^{*},R_1^{*}),(b_1^{*},R_2^{*})\}$ satisfying
\begin{equation*}
b_0^{*}=\pi_0R_1^{*},\quad b_1^{*}=R_2^{*}\quad\text{and}\quad R_1^{*}\leq\min\left\{\tfrac{\beta(1-\beta)}{\beta-\pi_0}b_1^{*},y_1+b_1^{*}\right\}
\end{equation*}
constitutes an equilibrium in which the strategic borrower defaults with probability one at both dates 1 and 2.\footnote{Evidently, the history at which the borrower defaults at $t=2$ is not reached in equilibrium.}
\item[(ii)] Any sequence of contracts $\{(b_0^{*},R_1^{*}),(b_1^{*},R_2^{*})\}$ satisfying
\begin{equation*}
b_0^{*}=\pi_0R_2^{*},\quad b_1^{*}=R_1^{*}\quad\text{and}\quad R_2^{*}\leq y_1+y_2
\end{equation*}
constitutes an equilibrium in which the strategic borrower defaults with probability
\begin{equation*}
\delta_1^{*}(b_1^{*},R_2^{*};\pi_0)=1-\tfrac{\pi_0}{1-\pi_0}\left(\tfrac{R_2^{*}}{b_1^{*}}-1\right)
\end{equation*}
at date 1 and defaults with probability one at date 2.
\end{enumerate}
\item[III.] If $\pi_0\geq\beta$, then in addition to the equilibria described in cases (I) and (II), there exists another class of equilibria with the sequence of contracts $\{(b_0^{*},R_1^{*}),(b_1^{*},R_2^{*})\}$ satisfying
\begin{equation*}
b_0^{*}=R_1^{*},\quad b_1^{*}=\pi_0R_2^{*},\quad R_1^{*}\leq y_1+b_1^{*}\quad\text{and}\quad R_2^{*}\leq y_2,
\end{equation*}
This corresponds to an equilibrium in which the strategic borrower never defaults at date $1$ (and always defaults at date 2).
\end{enumerate}
\end{lemma}

\begin{proof}
See \ref{proof:lemmasignal}.
\end{proof}

The equilibrium set expands as $\pi_0$ increases.

When the lenders' prior is very low ($\pi_0<\beta^2$), the honest type of borrower will not find it optimal to ask for a loan at either date 0 or 1, so the only equilibrium will be autarky. For intermediate values of $\pi_0$ (when $\beta^2\leq\pi_0<\beta$), there exist equilibria in which borrowing occurs on date 0, then on date 1 the honest borrower repays the loan, successfully separates from the strategic type and keeps on borrowing at a risk-free rate, while the strategic type defaults on the date-0 loan with probability one.

In addition, there exist equilibria in which the strategic type randomizes between defaulting and repaying at date 1; in this class of equilibria, the honest type can only partially separate himself from the strategic type by repaying the date-1 loan. Nevertheless, the lenders' posterior $\pi_1$ following loan repayment is sufficiently high to induce the honest type to ask for positive borrowing at the end of date 1.

Finally, when $\pi_0$ is very large, there also exist equilibria in which at date 1, the strategic type repays the date-0 loan $(b_0,R_1)$ with probability one, thus pooling with the honest type. This requires a large prior ($\pi_0\geq\beta$), because the honest type should be willing to borrow at date 1 despite the fact that repayment of the date-0 loan did not bring any `reputational reward' for him (since we have $\pi_1=\pi_0$ due to pooling).

All of those equilibria are supported by out-of-equilibrium beliefs, according to which the lenders attribute any deviation from the pooling contract to the strategic type. This restriction passes the refinement of \cite{ChoKreps1987}. To see this, first consider the borrower who deviates from the equilibrium contract at date 1. Now, the lenders know that if it were the strategic type, he would default with certainty at date 2 -- so, no one will provide the borrower the contract $(b_1,R_2)$ on the next date \emph{if the deviation from the equilibrium contract offered at the preceding date 0 has revealed that his type is strategic}. And if it happens so that at date 1, the borrower cannot repay the previous-period debt $R_1$ unless he can borrow $b_1$, the honest type will be \emph{less} willing to deviate from the equilibrium contract $(b_1,R_2)$, because this would force him into default on the previous-period loan, from which he would suffer an infinite utility loss. So, since the strategic borrower looses less by deviating from the equilibrium contract at date 1, the lenders' update upon observing the deviation passes the criterion.

On the other hand, at date 0, both types find it \emph{equally} unattractive to deviate to a contract revealing to the lenders that the borrower is of the strategic type: knowing the borrower's type, the date-1 lenders would deny him the loan. In turn, this implies that any positive repayment $R_1$ would induce default with probability one by the strategic type at date 1. Coupled with the date-0 lenders' zero-profit condition, this implies that the only contract to which the borrower can deviate \emph{and} which will be accepted by the lenders is $(b_0,R_1)=(0,0)$ -- which is equally unattractive to both types.\footnote{A special case of this unravelling argument would be used to prove that only autarkic outcome is feasible for the case when the borrower is commonly known to be strategic ($\pi_0=0$).}

To narrow down the set of possible equilibrium outcomes, in what follows, we will restrict ourselves to the \emph{undefeated equilibria} (see \cite{Mailathetal1993}), which is a refinement that selects the best equilibrium from the point of view of the honest type. Furthermore, the assumption that at date $t$, the borrower cannot commit to the contracts that he will offer at future dates requires us to impose \emph{sequential rationality} ensuring that the borrower behaves optimally at each history, irrespective of whether it is actually reached in equilibrium. As we will show in Proposition \ref{pardomcontr} below, the borrower will apply for the loan \emph{only if his initial reputation is large enough}, namely $\pi_0\geq\beta$.

\newcommand\grepfirst{%
\begin{tikzpicture}[scale=0.6]
\draw[thick, <->] (0,13) node[left]{$b_1$} -- (0,0) node[left]{0} -- (21,0) node[below]{$R_2$};
\draw[thin, dotted] (0,0) -- (11,11) node[above]{$45^{\circ}$};
\draw[thin, dotted] (0,0) -- (8,4);
\draw[thin, dotted] (0,4) node[left]{$\overline{R}_1$} -- (16,4);
\draw[ultra thick] (0,0) -- (4,4) -- (8,4) -- (16,8);
\draw[thick, dotted] (16,0) node[below]{$y_2$} -- (16,8);
\draw[thin] (2,0) arc (1:27:2) node[below right]{\tiny{$\pi_0$}};
\draw[dotted] (0,8) node[left]{$\pi_0y_2$} -- (16,8);
\draw[dotted] (4,0) node[below]{$\overline{R}_1$} -- (4,4);
\draw[dotted] (8,0) node[below]{${\overline{R}_1}/{\pi_0}$} -- (8,4);
\draw[ultra thick, dashed] (-2,2) -- (19,9) node[right]{$u_2^{*}$};
\draw[thin, dotted] (-2,2) --(1,2);
\draw[thin] (1,2) arc (1:20:3) node[below right]{\tiny{$\beta$}};
\draw [fill] (4,4) circle[radius=4pt];
\draw [fill] (8,4) circle[radius=4pt];
\draw [fill] (16,8) circle[radius=4pt];
\end{tikzpicture}%
}

\newcommand\grepsecond{%
\begin{tikzpicture}[scale=0.6]
\draw[thick, <->] (0,13) node[left]{$b_0$} -- (0,0) node[left]{0} -- (23,0) node[below]{$R_1$};
\draw[thin, dotted] (0,0) -- (11,11) node[above]{$45^{\circ}$};
\draw[thin, dotted] (0,0) -- (6,3);
\draw[thin, dotted] (4,0) node[below]{$\overline{R}_1$} -- (4,3);
\draw[ultra thick] (0,0) -- (4,4) -- (4,2) -- (16,8);
\draw[thick, dotted] (16,0) node[below]{$\overline{R}_1+y_1$} -- (16,8);
\draw[thin] (2,0) arc (1:27:2) node[below right]{\tiny{$\pi_0$}};
\draw[dotted] (0,8) node[left]{$\pi_0(\overline{R}_1+y_1)$} -- (16,8);
\draw [fill] (4,4) circle[radius=4pt] node[right]{pooling};
\draw [fill] (16,8) circle[radius=4pt] node[right]{separating};
\end{tikzpicture}%
}

\begin{figure}
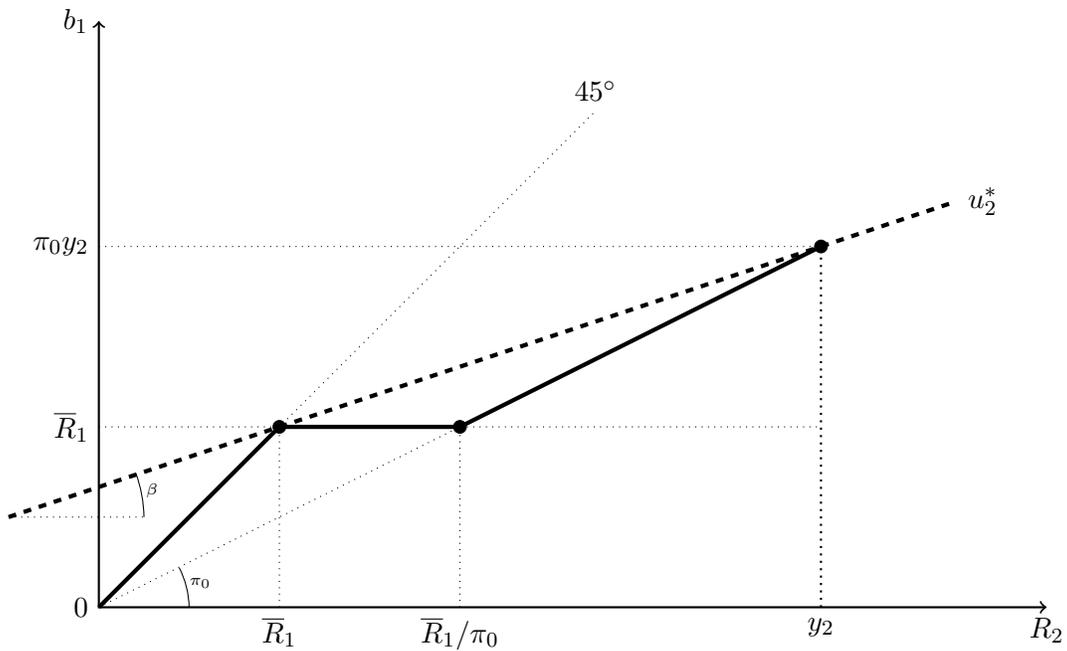
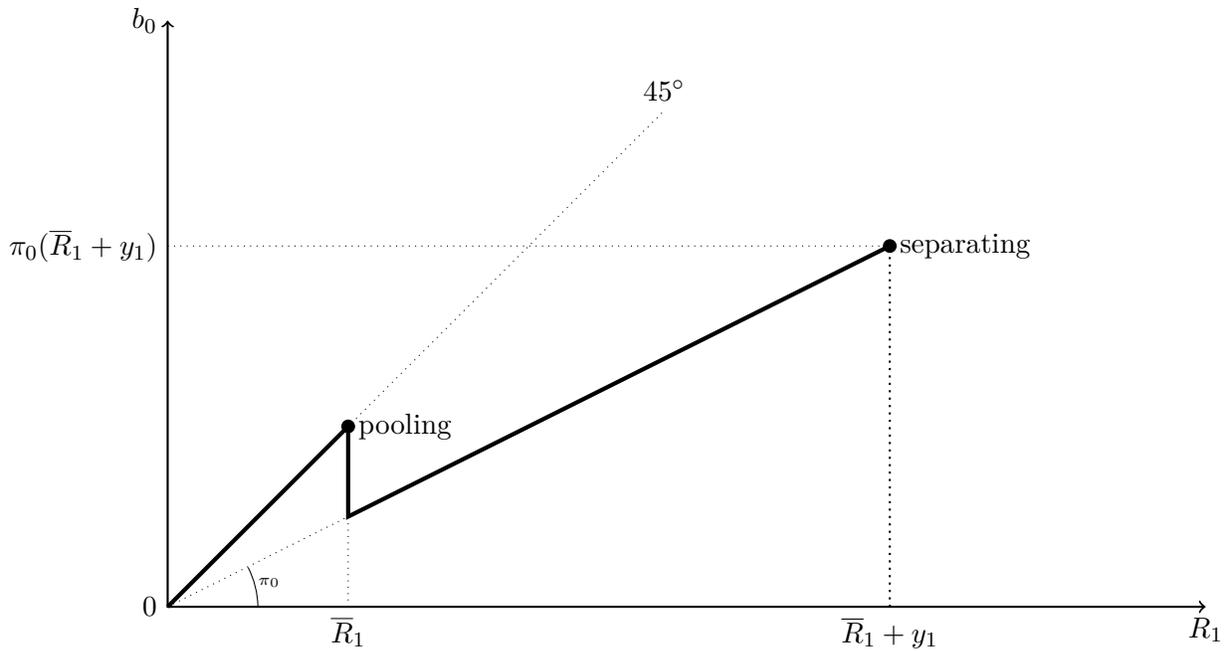

\centering
\subfloat[Determination of the maximal feasible riskless date-1 loan $\overline{R}_1$.]{\grepfirst}
\\
\subfloat[Feasible date-0 loan contracts, given $\overline{R}_1$.]{\grepsecond} 
\caption{The set of feasible lending contracts at $t=1,2$.}
\label{fig:feasdate}
\end{figure}

Figure \ref{fig:feasdate} illustrates the set of feasible lending contracts on dates $t=1$ and $t=0$, and the borrower's optimal choice is determined by adding the honest type's indifference curves. On both dates the borrower effectively faces a trade-off between applying for the small loan at a low interest rate and applying on a large loan at a high interest rate.

Figure \ref{fig:feasdate}(a) illustrates the set of date-1 lending contracts that satisfy the lenders' zero-profit condition. The additional restriction $R_2\leq y_2$ is the \emph{feasibility} condition: the honest type should not be forced into default at date 2, and hence his income $y_2$ should not be smaller than the amount promised for repayment, $R_2$.

To see how this graph is constructed, consider the situation at date $t=1$. Fix $R_1$, the repayment on the date-0 loan (which was determined at the previous date $t=0$) and consider the various regions for the date-1 borrowing level $b_1$.

First suppose that at date 1, the borrower applies for the contract $(b_1,R_2)$ such that the amount borrowed at this date is less than the repayment on the previous-period loan: $b_1<R_1$. In order to be able to borrow, the individual has to repay the previous-period debt, $R_1$. The strategic borrower could, in principle, repay $R_1$, borrow $b_1$ and default on this loan on the next date. However, this will bring him the net gain of $b_1-R_1<0$. Therefore, for $b_1<R_1$, the strategic borrower will be better off \emph{defaulting} on $R_1$. So, observing the borrower who has repaid the loan $R_1$, the date-1 lenders infer that the borrower is \emph{honest}. Accordingly, the posterior belief is updated to $\pi_1=1$, and thus will allow him to borrow at a risk-free rate ($b_1=R_2$). Therefore, for $b_1<R_1$, the segment of the zero-profit condition coincides with the 45 degree line starting from the origin.

Now suppose instead that at $t=1$, the borrower picks the loan contract for which $b_1>R_1$. In this case, by repaying the previous-period debt $R_1$, borrowing $b_1$ and defaulting on this debt at date 2, the strategic borrower would get a net gain of $b_1-R_1>0$. Therefore, he will strictly prefer to repay $R_1$. Correspondingly, the date-1 lender who sees that the borrower has repaid $R_1$ and is now asking for a loan $b_1>R_1$ will not learn anything from the fact of repayment. His posterior belief that the lender is honest will stay unchanged ($\pi_1=\pi_0$), so the upper segment of the zero-profit condition corresponds to the line that starts at the origin and that has the slope of $\pi_0$.

Finally, when $b_1=R_1$, the contract allows exact roll-over of the previous-period debt, and the strategic borrower is indifferent between defaulting and repaying, since his net gain from repayment is zero ($b_1-R_1=0$). As we move along the segment and $R_2$ increases from $R_1$ to $R_1/\pi_0$, the probability with which the strategic type defaults on $R_1$ gradually decreases from one to zero.

Since the strategic type always mimics the honest type, in the undefeated equilibrium, the borrower picks the contract that is optimal for the honest type. The dashed line depicts the honest borrower's indifference curve. The contract $(b_1,R_2)$ brings him the net increase in utility of $b_1-\beta R_2$ -- therefore, the honest type's indifference curve is represented by the straight line with the positive slope equal to $\beta$. The borrower's utility increases in the north-west direction.

Depending on the values of $\beta$ and $\pi_0$, two generic cases for the borrower's optima might arise: if $\beta$ is relatively high and $\pi_0$ is relatively low, he would pick the contract $(b_1,R_2)=(R_1,R_1)$, whereas if $\beta$ is relatively low and $\pi_0$ is relatively high, he would prefer the contract $(b_1,R_2)=(\pi_0y_2,y_2)$. Figure \ref{fig:feasdate}(a) represents the case when the honest borrower is exactly indifferent between the two contracts. Let us define by $\overline{R}_1$ the maximal feasible riskless loan that the honest type can make at date $t=1$. The expression for $\overline{R}_1$ is given by\footnote{Derivation is given in the proof of Proposition \ref{pardomcontr} in the \ref{proof:pardomcontr}.}
\begin{equation}
\overline{R}_1=\max\left\{\frac{(\pi_0-\beta)y_2}{1-\beta},0\right\}.
\end{equation}

As we can see, $\overline{R}_1$ is positive only provided that $\pi_0\geq\beta$. By contrast, whenever $\pi_0<\beta$, then irrespective of the promised repayment $R_1$, the date-0 lenders will recognize that at the next date, the honest type will prefer the contract that exactly rolls over the debt from $t=1$ to $t=2$, that is, $b_1=R_2$. But the strategic type will prefer to default on this contract, and hence the loan with $R_1>\overline{R}_1=0$ will not be riskless. For $\pi_0>\beta$, $\overline{R}_1$ stands for the amount that would make him indifferent between the two contracts at date 2.

Figure \ref{fig:feasdate}(b) represents the set of date-0 feasible contracts satisfying the lenders' zero-profit condition, for a given anticipated continuation contract at date 1. First, suppose he were to offer a loan contract $(b_0,R_1)$ with $R_1<\overline{R}_1$. In that case, both types of borrower will repay the loan at date 1 -- therefore, from the perspective of the date-0 lender, this loan is safe: we have $b_0=R_1$, and the zero-profit condition coincides with the 45 degree line. Next, suppose that $R_1>\overline{R}_1$. In that case, the date-0 lenders will expect the strategic borrower to default on the date-0 debt, and from the lenders' zero-profit condition, we will have $b_0=\pi_0R_1$. This corresponds to the line segment with the slope $\pi_0$. As we move down along the vertical segment, the strategic borrower's default probability gradually increases from zero to one.

It turns out that the honest borrower's choice whether (i) to pick the riskless contract and pool with the strategic type at date 1 (in order to repay the date-0 debt and subsequently pick the risky contract at date 1 and separate at date 2), or else (ii) pick the risky contract at date 0 and separate at date 1 depends on the growth of the borrower's non-financial income: specifically, define
\begin{equation}
g=\frac{y_2-y_1}{y_1}.
\end{equation}

As we claim, whenever $g>0$, there exists an intermediate region for $\pi_0$, such that the borrower prefers pooling at date 1. For higher values of $\pi_0$, early separation is preferable. The next proposition characterizes the unique undefeated equilibrium (the one that attains the maximal payoff for the honest type) for various levels of $\pi_0$:

\begin{proposition}
\label{pardomcontr}
The unique undefeated equilibrium is given by:
\begin{enumerate}
\item Autarky for $\pi_0<\beta$:
\begin{equation}
(b_0^{*},R_1^{*})=(0,0)\quad\text{and}\quad(b_1^{*},R_2^{*})=(0,0).
\end{equation}
\item Pooling at date 1 and separation at date 2 for $\beta\leq\pi_0<1-\frac{1-\beta}{1+g}$:
\begin{equation}
(b_0^{*},R_1^{*})=(\overline{R}_1,\overline{R}_1)\quad\text{and}\quad(b_1^{*},R_2^{*})=(\pi_0y_2,y_2).
\end{equation}
\item Separation at date 1 for $\pi_0\geq1-\frac{1-\beta}{1+g}$:
\begin{equation}
(b_0^{*},R_1^{*})=(\pi_0(\overline{R}_1+y_1),(\overline{R}_1+y_1))\quad\text{and}\quad(b_1^{*},R_2^{*})=(\overline{R}_1,\overline{R}_1).
\end{equation}
\end{enumerate}
The range of $\pi_0$ for which pooling is preferable is non-empty only for $g>0$.
\end{proposition}

\begin{proof}
See \ref{proof:pardomcontr}.
\end{proof}

Let us briefly summarize the main takeaway from this benchmark case. First, when $\pi_0<\beta$, the honest type will abstain from borrowing even if repayment of the loan at date 1 could separate him from the strategic type -- in that case, the only equilibrium will correspond to autarky. Second, for intermediate values of $\pi_0$, the honest borrower would prefer to pool with the strategic type at an early date and to separate only at a later date, whereas for very high initial reputation $\pi_0$, early separation would be preferable. Third, the range of $\beta$ and $\pi_0$ for which pooling is preferable expands when the income growth $g$ becomes larger.

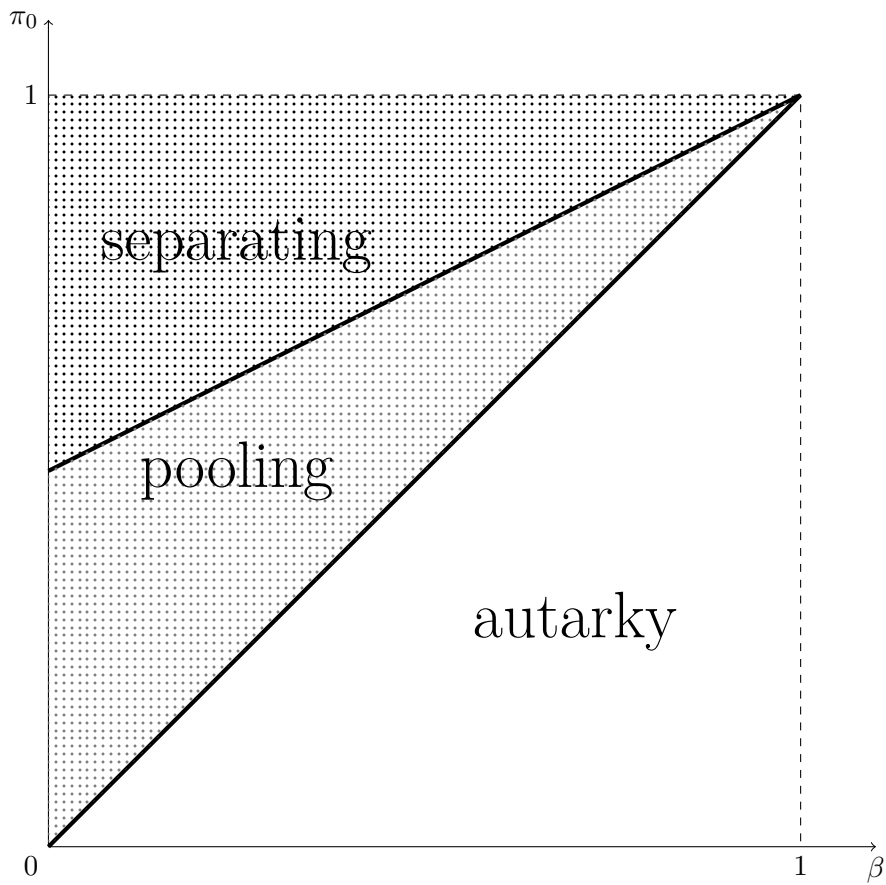
\begin{figure}[t]
\begin{center}
\begin{tikzpicture}[scale=10]
\draw[<->] (0,1.1) node[left]{$\pi_0$} -- (0,0) node[below left]{0} -- (1.1,0) node[below]{$\beta$};
\draw [ultra thick, domain=0:1] plot(\x,{\x});
\draw [ultra thick, domain=0:1] plot(\x,{(\x+1)/2});
\draw [dashed] (0,1) node[left]{1} -- (1,1) -- (1,0) node[below]{1};
\fill [pattern color=black, pattern=dots] (0,0.5) -- (0,1) -- (1,1) -- cycle;
\fill [pattern color=gray, pattern=dots] (0,0) -- (0,0.5) -- (1,1) -- cycle;
\node at (0.25,0.8){\Huge{separating}};
\node at (0.25,0.5){\Huge{pooling}};
\node at (0.7,0.3){\Huge{autarky}};
\end{tikzpicture}
\caption{Equilibrium outcomes for different $(\beta,\pi_0)$: case when $g>0$.}
\label{fig:eqrange}
\end{center}
\end{figure}

The parameter range for $(\beta,\pi_0)$ for which the undefeated equilibrium is characterized by autarky, pooling and separation is represented on Figure \ref{fig:eqrange}. Intuitively, if the borrower expects higher income growth ($y_2$ much higher than $y_1$), this increasing his maximal feasible riskless loan $\overline{R}_1$, making the pooling contract more attractive. On the other hand, the higher is the initial reputation $\pi_0$, the more tempting it is to separate early, because the adverse-selection discount $(1-\pi_0)$ decreases with $\pi_0$.

To conclude this section, we would like to stress once again that the date-1 borrowing constraint implicit in $\overline{R}_1$ is due to the borrower's inability to commit to future contracts. To see this most clearly, suppose we had $y_1=y_2=y$ (so that $g=0$) and let $\beta^2<\pi_0<\beta$. Proposition \ref{pardomcontr} tells us that at date 1, the borrower will pick an autarkic contract and obtain utility $\beta(1+\beta)y$. By contrast, suppose that at date 0, the borrower could \emph{commit} to picking the contract $(b_1,R_2)=(\pi_0y,y)$ at date 1. Then for any $R\leq\pi_0y$, the riskless contract $(b_0,R_1)=(R,R)$ would be feasible for him at date 0: the borrower's utility from the pair of contracts $(b_0,R_1)=(\pi_0y,\pi_0y)$ and $(b_1,R_2)=(\pi_0y,y)$ will be equal to $(\pi_0+\beta)y$, which is greater than the utility from autarky, since $\pi_0>\beta^2$.

As we will show in the next section, the possibility to use physical collateral may help the borrower overcome this commitment problem by creating the possibility of the adverse asset-price shock at date $t=1$ which will enable the honest type to lever up his reputation, thereby creating the additional incentive to hold the asset rather than selling it at the outset.

\section{\label{assetreputation}Physical collateral and reputation}

Now we introduce the possibility for the borrowers \emph{both} to pledge physical collateral and to build reputation: we consider the case when $\kappa_t\in\{0,1\}$ and $\pi_0>0$. Throughout the analysis we will impose the restriction that $\pi_0<\beta$. As we established in Proposition \ref{pardomcontr}, in the absence of the possibility to pledge the asset, reputation will not work: in both periods, the honest type will find it optimal to choose autarky. Furthermore, to facilitate exposition, we restrict our attention to the case when the income growth is zero, $g=\frac{y_2-y_1}{y_1}=0$, and thus we take equal non-financial incomes in both periods: $y_1=y_2=y$.

We show that the option to pledge the asset as collateral increases the value of the asset when the borrower can develop reputation. As suggested by the analysis in Section \ref{onlyasset}, the agent would prefer to sell his asset in period 0 at any price if he could not develop reputation for honesty.

We consider an equilibrium with the following structure:
\begin{enumerate}
\item At date $t=0$, there exist two threshold $\underline{p}_0$ and $\overline{p}_0$, such that
\begin{enumerate}
\item Both types keep the asset and pledge it as collateral $p_0\in[\underline{p}_0,\overline{p}_0]$;
\item Both types sell the asset when either $p_0<\underline{p}_0$ or $p_0>\overline{p}_0$.
\end{enumerate}

\item For the given date-0 loan contract $(b_0,R_1)$, in the date-1 continuation game when both types keep the asset, there exist two thresholds, $\underline{p}_1$ and $\overline{p}_1$, such that
\begin{enumerate}
\item Strategic type repays with probability one for $p>\overline{p}_1$;
\item Strategic type randomizes between default and repayment for $p_1\in[\underline{p}_1,\overline{p}_1]$;
\item Strategic type defaults with probability one for $p_1<\underline{p}_1$.
\end{enumerate}
\end{enumerate}

Notice that the types pool on their date-0 selling decisions.\footnote{To see that this is the only possible outcome, by contradiction suppose that for some price $p_0$, one type of borrower sells the asset while the other keeps it. If it is the honest borrower who sells the asset, whereas the strategic borrower keeps it, then the mere fact of offering the contract reveals that he is strategic. We know that the maximal amount that the strategic borrower will be able to borrow at date 0 is equal to $\E[p_1|p_0]=p_0$, the payoff which he would get by selling the asset.

Alternatively, suppose that at a price $p_0$, only the honest type keeps the asset. Since the strategic type has an option to default on the loan at date 1, for him the value of the asset is at least as large as the one for the honest type -- therefore, he will prefer to deviate and pool with the honest type.

Therefore, the selling decision cannot separate the two types.}

On the other hand, at date 2, the two types of borrowers pool for some realizations of $p_1$ and separate for other realizations: there is complete separation for $p_1<\underline{p}_1$, partial separation for $\underline{p}_1\leq p_1\leq\overline{p}_1$, and pooling for $p_1>\overline{p}_1$.

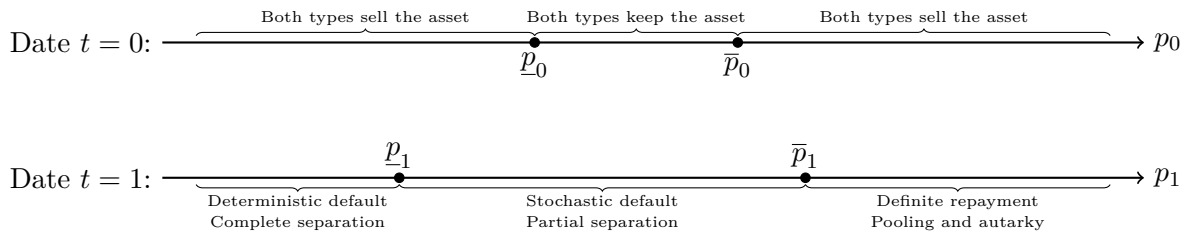
\begin{figure}
\begin{tikzpicture}[scale=0.9]
\draw [thick, ->] (-1.5,0) node[left]{Date $t=0$:} -- (13,0) node[right]{$p_0$};
\draw [decorate, decoration={brace}] (4,0.1) -- node[above] {\tiny{Both types keep the asset}} (7,0.1);
\draw [decorate, decoration={brace}] (7,0.1) -- node[above] {\tiny{Both types sell the asset}} (12.5,0.1);
\draw [decorate, decoration={brace}] (-1,0.1) -- node[above] {\tiny{Both types sell the asset}} (4,0.1);
\draw [fill] (7,0) circle[radius=2pt] node[below]{$\overline{p}_0$};
\draw [fill] (4,0) circle[radius=2pt] node[below]{$\underline{p}_0$};
\draw [thick, ->] (-1.5,-2) node[left]{Date $t=1$:} -- (13,-2) node[right]{$p_1$};
\draw [decorate, decoration={brace}] (12.5,-2.1) -- node[below] {\tiny{Definite repayment}} (8,-2.1);
\node[below] at (10.25,-2.41){\tiny{Pooling and autarky}};
\draw [decorate, decoration={brace}] (8,-2.1) -- node[below] {\tiny{Stochastic default}} (2,-2.1);
\node[below] at (5,-2.41){\tiny{Partial separation}};
\draw [decorate, decoration={brace}] (2,-2.1) -- node[below] {\tiny{Deterministic default}} (-1,-2.1);
\node[below] at (0.5,-2.41){\tiny{Complete separation}};
\draw [fill] (2,-2) circle[radius=2pt] node[above]{$\underline{p}_1$};
\draw [fill] (8,-2) circle[radius=2pt] node[above]{$\overline{p}_1$};
\end{tikzpicture}
\caption{The structure of equilibrium with both collateral and reputation.}
\label{eqstruc}
\end{figure}

The structure of equilibrium is schematically illustrated on Figure \ref{eqstruc}.

There are several novel elements in comparison to the two benchmark cases analyzed in Sections \ref{onlyasset} and \ref{onlyreputation}. First, for the bottom tail of asset prices, at the interim date $t=1$, the types can be separated via default. Second, even in the undefeated equilibrium (in which the honest type proposes the contract that maximize his utility), the strategic borrower can randomize between default and repayment for intermediate price range.

Construction of this equilibrium proceeds as follows. We start by analyzing continuation game at date 1 that follows borrowing at date 0 and repaying the debt at date 1. For a given lenders' posterior $\pi_1$, we characterize the optimal contract $(b_1^{*},R_2^{*})$ that is proposed by the honest type. Second, we compare the honest type's utility from selling the asset to the one from keeping it (in order to borrow until date 2), and we show that selling the asset always dominates keeping it. Third, we derive the lenders' posterior $\pi_1$ from the strategic borrower's indifference condition, and subsequently determine the default probability $\delta_1^{*}(p_1;R_1)$ and the two thresholds $\underline{p}_1$ and $\overline{p}_1$.

Then we proceed to characterizing optimal loan contract $(b_0,R_1)$. We show that, as in Section \ref{onlyasset}, the borrower's problem has bang-bang solution: the best non-autarkic debt contract is such that the strategic type defaults with probability one at date 1.

Finally, we characterize the borrower's selling decision at date 0. We show that both types of borrower keep the asset for $p_0\in[\underline{p}_0,\overline{p}_0]$.

\subsection{Borrowing and default decisions at date 1}

First, take the continuation game when the borrower has kept the asset and repaid the date-0 debt. The next lemma shows that, irrespective of the realization of $p_1$, the borrowers always finds it optimal to sell the asset at $t=1$.

\begin{lemma}
\label{alwayssell}
At date $t=1$, the borrower always chooses to sell the asset.
\end{lemma}

\begin{proof}
See \ref{proof:alwayssell}.
\end{proof}

Intuitively, the asset which yields no dividends is valuable to the borrower only to the extent that it enables him to lever up his reputation: when the realization of $p_1$ is exceptionally low, the strategic type finds it most tempting to default, and thus repayment of the previous-period loan allows the honest type to separate from the strategic type.

The strategic type will indeed find it optimal to default when the price is low and to repay when the price is high. The next proposition gives a formal characterization of the default decision.

\begin{proposition}
\label{initdef}
Given $R_1$, the strategic type always defaults when $p_1<\underline{p}_1$, always repays when $p_1>\overline{p}_1$, and randomizes between defaulting and repaying whenever $p_1\in[\underline{p}_1,\overline{p}_1]$, where
\begin{equation}
\underline{p}_1=R_1-y\quad\text{and}\quad\overline{p}_1=R_1.
\end{equation}
\end{proposition}

\begin{proof}
See \ref{proof:initdef}.
\end{proof}

\newcommand\lendpost{%
\begin{tikzpicture}[scale=0.8]
\draw[thick, <->] (0,10) node[left]{$\pi_1$} -- (0,0) node[left]{0} -- (15,0) node[below]{$p_1$};
\draw[dotted] (3,0) node[below]{$R_1-y$} -- (3,5);
\draw[dotted] (6,0) node[below]{$R_1-\beta y$} -- (6,3);
\draw[dotted] (10,0) node[below]{$R_1$} -- (10,1);
\draw[dotted] (0,3) node[left]{$\beta$} -- (6,3);
\draw[dotted] (0,1) node[left]{$\pi_0$} -- (10,1);
\draw[dashed] (10,1) -- (10,3);
\draw[ultra thick] (0,5) node[left]{1} -- (3,5) --(6,3) -- (10,3);
\draw[ultra thick] (10,1) -- (14,1);
\draw [decorate, decoration={brace}] (0,0.1) -- node[above] {\tiny{Complete separation}} (3,0.1);
\draw [decorate, decoration={brace}] (3,0.1) -- node[above] {\tiny{Partial separation}} (6,0.1);
\draw [decorate, decoration={brace}] (6,0.1) -- node[above] {\tiny{Credit rationing}} (10,0.1);
\draw [decorate, decoration={brace}] (10,0.1) -- node[above] {\tiny{Pooling and autarky}} (14,0.1);
\draw [fill] (10,3) circle[radius=2pt];
\draw [fill=white] (10,1) circle[radius=2pt];
\end{tikzpicture}%
}

\newcommand\lendrand{%
\begin{tikzpicture}[scale=0.8]
\draw[thick, <->] (0,10) node[left]{$\delta_1$} -- (0,0) node[left]{0} -- (17,0) node[below]{$p_1$};
\draw[dotted] (3,0) node[below]{$R_1-y$} -- (3,5);
\draw[dotted] (6,0) node[below]{$R_1-\beta y$} -- (6,65/16);
\draw[dotted] (0,65/16) node[left]{$\frac{\beta-\pi_0}{\beta(1-\pi_0)}$} -- (6,65/16);
\draw[dashed] (10,0) node[below]{$R_1$} -- (10,65/16);
\draw [ultra thick] (0,5) node[left]{1} -- (3,5);
\draw [ultra thick, domain=3:6] plot(\x,{5-1.25*(7/(10-\x)-1)});
\draw [ultra thick] (6,65/16) -- (10,65/16);
\draw [ultra thick] (10,0) -- (16,0);
\draw [fill] (10,65/16) circle[radius=2pt];
\draw [fill=white] (10,0) circle[radius=2pt];
\draw [decorate, decoration={brace}] (0,0.1) -- node[above] {\tiny{Deterministic default}} (3,0.1);
\draw [decorate, decoration={brace}] (3,0.1) -- node[above] {\tiny{Stochastic default}} (6,0.1);
\draw [decorate, decoration={brace}] (6,0.1) -- node[above] {\tiny{Stochastic lending}} (10,0.1);
\draw [decorate, decoration={brace}] (10,0.1) -- node[above] {\tiny{Repayment and selling}} (16,0.1);
\end{tikzpicture}%
}

\begin{figure}
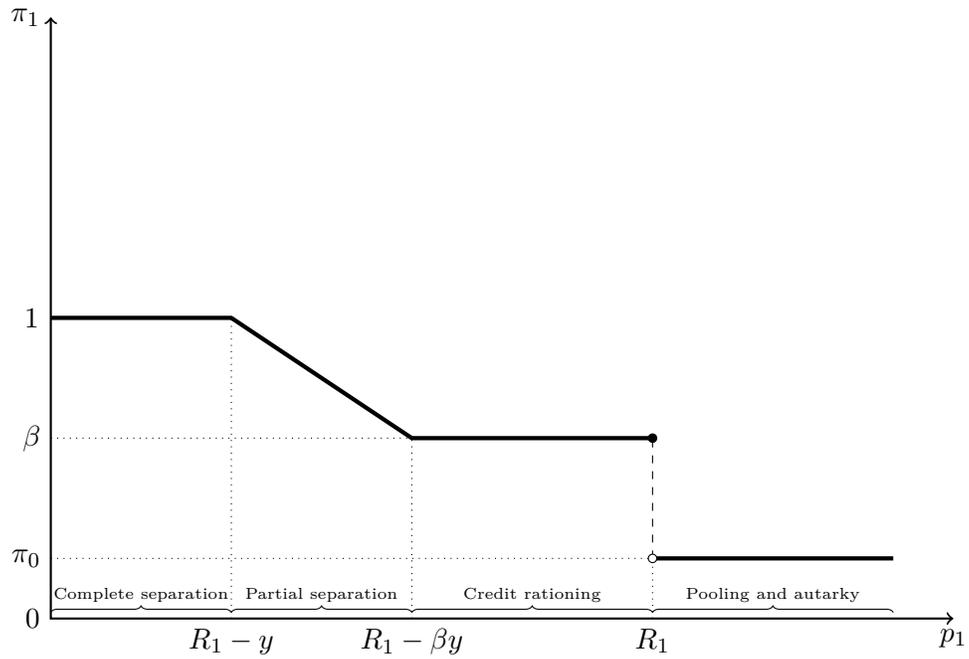
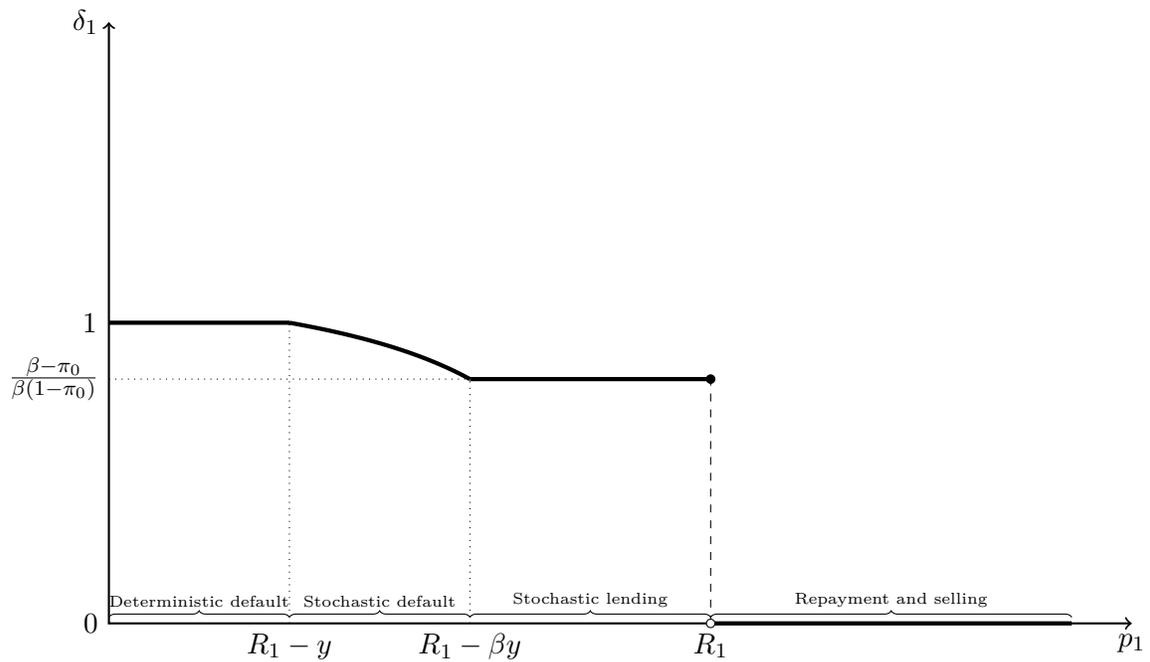

\centering
\subfloat[Lenders' posterior as a function of $p_1$, for given $R_1$.]{\lendpost}
\\
\subfloat[Strategic borrower's default probability as a function of $p_1$, for given $R_1$.]{\lendrand} 
\caption{Default incentives and reputation as a function of $p_1$.}
\label{fig:lendpost}
\end{figure}

The lenders' posterior is depicted on Figure \ref{fig:lendpost}(a). For $p_1>\overline{p}_1$, debt repayment does not lead to the lenders' posterior update, so that we have $\pi_1=\pi_0$. This happens due to the fact that for $p_1>R_1$, the strategic borrower is better off repaying his date-0 debt and realizing his option to sell the asset. Therefore, the types pool at very high prices, and for the restriction on the lenders' prior that we focus on ($\pi_0<\beta$), the honest type prefers to stay in autarky.

Along the same lines, for extremely low realizations of $p_1$, repayment indicates that the type is honest and raises the lenders' posterior to $\pi_1=1$. In particular, whenever $p_1+y<R_1$, the costs of loan repayment exceed the maximal possible gain for the strategic borrower. As $p_1$ increases from $\underline{p}_1$ to $\overline{p}_1$, the lenders' posterior decreases from one to $\pi_0$.

To gain intuition concerning the jump at $p_1=R_1$, observe that there cannot exist an equilibrium with $\pi_1\in(\pi_0,\beta)$: as we showed in the previous section, the honest type will not continue borrowing at date 1 unless $\pi_1\geq\beta$. Therefore, in case if the strategic type were to repay the loan, the lenders would update their beliefs to $\pi_1=1$. On the other hand, the strategic type will find such a deviation profitable for $p_1$ close to $R_1$. At the same time, in order to keep the honest type indifferent between borrowing and staying in autarky, the lenders' posterior should remain at $\pi_1=\beta$. However, since the strategic type would \emph{strictly} prefer to apply for the loan if he could obtain credit with certainty, in order to make the strategic type indifferent between defaulting and repaying, loan provision should be stochastic: the probability that the lender accepts the contract should be less than one.

Finally, Bayes' rule for the update of $\pi_1$ can be derived from the strategic borrower's indifference condition. The lenders' posterior belief $\pi_1(p_1;R_1)$ is depicted on Figure \ref{fig:lendpost}(a) and the strategic type's default probability $\delta_1^{*}(p_1;R_1)$ is depicted on Figure \ref{fig:lendpost}(b).

\subsection{Selling decision at date $t=0$}

Having characterized equilibrium in the continuation game after the borrower has kept the asset, we will now determine for which set of $p_0$ the honest (and thus, also the strategic) type will prefer to keep the asset for borrowing.

This is stated in the next theorem.

\begin{theorem}
\label{mainresult}
There exists a lower bound for $\pi_0$ such that the honest type finds it profitable to keep the asset and borrow at date $0$ for any $\pi_0\geq\pi_0^{*}(\beta,y,p_0)$. The lower bound is given by
\begin{equation}
\label{eq:suffcond}
\pi_0^{*}(\beta,y,p_0)=\max\left\{1-(1-\beta)\left(\frac{p_0}{y}+\frac{\beta}{2}\right),\frac{4(1-\beta)}{(5-\beta)}\left(\frac{p_0}{y}-2\right)\frac{p_0}{y}+\frac{\beta(3+\beta^2)-4(1+\beta^2)}{5-\beta}\right\}.
\end{equation}

Furthermore, this lower bound is less tight than in the no-asset case, that is, $\pi_0^{*}(\beta,y,p_0)<\beta$, whenever
\begin{equation}
\label{eq:bounds}
\frac{p_0}{y}\in\left(1-\frac{\beta}{2},1+\sqrt{\left(\frac{2}{1-\beta}-\frac{\beta}{2}\left(1-\frac{\beta}{2}\right)\right)}\right).
\end{equation}
\end{theorem}

\begin{proof}
See \ref{proof:mainresult}.
\end{proof}

This result characterizes the bound $\pi_0^{*}(\beta,y,p_0)$ for the borrower's initial reputation, above which he would prefer to keep the asset and use it as collateral. This bound is less tight than $\beta$ (the one that was derived in Proposition \ref{pardomcontr} in Section \ref{onlyreputation}), whenever the ratio of the borrower's financial to non-financial income $\frac{p_0}{y}$ is neither too large nor too small.

Theorem \ref{mainresult} points out the ingredients that facilitate reputation building.

First, there should be significant gains from borrowing; in the context of this model, this implies that the discount factor has to be low: indeed, the reduction in $\beta$ expands the set of $y$ and $p_0$ for which the condition \eqref{eq:suffcond} is satisfied. Second, the date-0 asset price must be neither too high nor too low. When $p_0$ is very high, for most of realizations for $p_1$, the strategic borrower repays the date-0 loan, which in turn does not allow complete separation of types, and thus wipes out gains from continued borrowing for the honest type; at the same time, the individual's borrowing capacity is directly affected by $p_0$, since this is how much (in expectation) the lenders could recover if the strategic type were to default with probability one -- therefore, when $p_0$ is very low, borrowing becomes so expensive that the honest type would rather prefer to sell the asset and remain in autarky. The date-0 loan contract proposed by the honest type optimally trades off the benefits from separating against the costs of the loan: $R_1$ is made such that in the date-1 continuation game, the strategic borrower defaults for low $p_1$, repays for high $p_1$, and randomizes for intermediate $p_1$.

To gain intuition for this result, it should be kept in mind that the strategic borrower's behavior is still determined by optimizing behavior of the honest type: that is, some continuation games are characterized by pooling equilibria. This stands in sharp contrast with the result presented in section \ref{onlyasset}, in which there is no scope for mimicking, because the borrower's type is common knowledge.\footnote{In particular, as is evident from Proposition \ref{prop:borrow}, the borrower prefers to sell the asset in period $t=0$ and remain in autarky.}

\section{\label{robustness}Discussion and extensions}

This section discusses the assumptions and implications of the model developed in section \ref{assetreputation}. Then we will give a brief account of several extensions.

First, we consider the case when the asset delivers the borrower a per-period \emph{non-pledgeable} income $x>0$. The insights from this modification lie half-way between our analysis in Sections \ref{onlyasset} and \ref{assetreputation}: unlike in the only-asset case, the borrower need not prefer to sell the asset immediately, because the prospect of losing $x$ provides an additional incentive to repay, which creates mutual gains from trade. However, the comparison of Figures \ref{eqstruc} and \ref{eqstruccollat} reveals the difference: when $p_0$ is very low, the possibilities to build reputation are limited,\footnote{Since the asset price is bounded below, it cannot fall by much when it is already close to zero.} and thus in the model of Section \ref{assetreputation}, the borrower would prefer to sell the asset. By contrast, if the per-period income $x$ delivered by the asset is exogenous, the benefits of keeping it are independent of its price, which modifies the borrower's behavior at dates 0 and 1.

Second, we give a brief account of the generalized finite-horizon version with the representative borrower, in which the number of periods $T$ tends to infinity and give an informal discussion of how the equilibrium in the three-period setup may be adapted to that context.

Third, we take up an overlapping generations' perspective: the time horizon will be infinite, and in each period $t\geq2$, there will exist three generations of borrowers: young, middle-aged, and old. In each generation, the proportion of honest types within each young generation will be equal to $\pi_0$. At each date $t$, all the generations will face the same asset price $p_t$. This extension will allow us to untie the impact of asset price fluctuations, on the one hand, from the borrowers' life cycle considerations.

Finally, we address robustness and possible generalizations.

\subsection{Borrowing against the asset that delivers the stream of dividends}

In this subsection, we generalize the setup of section \ref{onlyasset} by allowing the asset to yield a per-period dividend stream of $x>0$. This sum $x$ represents \emph{non-pledgeable income}, meaning that it accrues exclusively to the borrower, and thus if the lender seizes the asset, she cannot get a claim on $x$.\footnote{This assumption that the income $x$ accrues privately to the borrower and is lost whenever there is a default may be motivated as follows: for instance, as in the search literature, the agent's private value for holding the asset can be driven by his hedging needs: in case if short-selling would require the individual to search for the asset lender and/or bargain over the lending fee, the borrower may attribute a higher valuation for the given stream of dividends than the lenders. See \cite*{Duffieetal2002}, \cite{Duffieetal2007}.} We will see how such an assumption will modify the optimal contract in case when the entrepreneur can borrow only against collateral.

Since the lenders know with certainty that the borrower is of the `strategic' type, so that $\pi_0=0$, loan repayment cannot be enforced by reputation, the lenders will accept only the collateralized loan contracts -- those with $\kappa_t=1$.\footnote{The only feasible loan contract without collateral is the `autarkic' contract $\mathcal{B}_t=(0,0,0)$.} Hence, the agent who has sold the asset can no longer borrow.

The next proposition characterizes the individual's optimal selling, borrowing and default decisions.

\begin{proposition}
\label{prop:borrowasset}
At date $0$, the individual sells the asset if and only if the asset price exceeds the threshold: $p_0\geq\hat{p}_0$, where
\begin{equation}
\label{date0threshasset}
\hat{p}_0=\tfrac{2x}{1-\beta^2}.
\end{equation}
Otherwise, an individual keeps the asset and offers the contract
\begin{equation}
(\hat{b}_0,\hat{R}_1)=\left(2x,2x\right).
\end{equation}

At date $1$, the individual who kept the asset (and borrowed against it) repays his loan and offers the loan contract
\begin{equation}
(\hat{b}_1,\hat{R}_2)=
\begin{cases}
(x,x)&\text{if }p_1<\hat{p}_1,\\
\left(p_1+\frac{1-2\beta}{p_1}\left(\frac{x}{2(1-\beta)}\right)^2,2p_1-\frac{\beta x}{1-\beta}\right)&\text{if }p_1\geq\hat{p}_1,
\end{cases}
\end{equation}
where
\begin{equation}
\hat{p}_1=\frac{x}{2(1-\beta)}.
\end{equation}

At date $2$, the individual defaults on the date-$1$ loan whenever the asset price falls below the threshold: $p_2<\hat{p}_2$, where
\begin{equation}
\hat{p}_2=R_2-x.
\end{equation}
\end{proposition}

\begin{proof}
See \ref{proof:borrowasset}.
\end{proof}

\begin{figure}
\begin{center}
\begin{tikzpicture}[scale=0.7]
\draw [thick, ->] (-1.5,0) node[left]{Date $t=0$:} -- (13,0) node[right]{$p_0$};
\draw [decorate, decoration={brace}] (12.5,-0.1) -- node[below] {Selling} (6,-0.1);
\draw [decorate, decoration={brace}] (6,-0.1) -- node[below] {Keeping} (-1,-0.1);
\draw [fill] (6,0) circle[radius=2pt] node[above]{$\hat{p}_0$};
\draw [thick, ->] (-1.5,-2) node[left]{Date $t=1$:} -- (13,-2) node[right]{$p_1$};
\draw [decorate, decoration={brace}] (12.5,-2.1) -- node[below] {Risky contract} (8,-2.1);
\draw [decorate, decoration={brace}] (8,-2.1) -- node[below] {Safe contract} (-1,-2.1);
\draw [fill] (8,-2) circle[radius=2pt] node[above]{$\hat{p}_1$};
\draw [thick, ->] (-1.5,-4) node[left]{Date $t=2$:} -- (13,-4) node[right]{$p_2$};
\draw [decorate, decoration={brace}] (12.5,-4.1) -- node[below] {Repayment} (4,-4.1);
\draw [decorate, decoration={brace}] (4,-4.1) -- node[below] {Default} (-1,-4.1);
\draw [fill] (4,-4) circle[radius=2pt] node[above]{$\hat{p}_2$};
\end{tikzpicture}
\end{center}
\caption{The structure of equilibrium with collateral only.}
\label{eqstruccollat}
\end{figure}
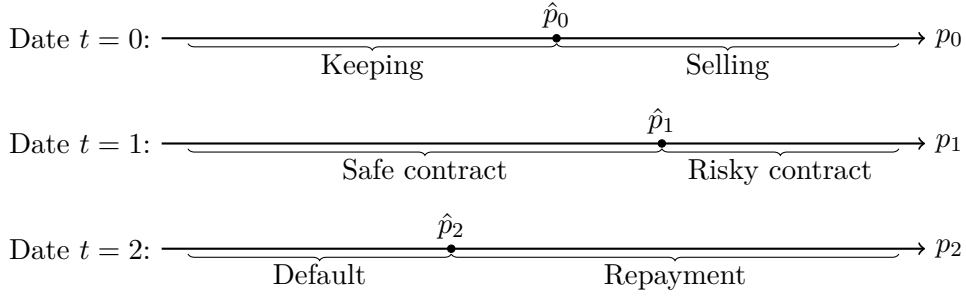

The structure of equilibrium is illustrated on Figure \ref{eqstruccollat}.

At this point it should be observed that the individual who borrowed at date $0$ never defaults on the loan at date $1$ and never sells the asset at dates $1$ and $2$.

To gain intuition for the last claim, notice that selling the asset is weakly dominated by borrowing against it. This result does not depend on the specificities of the linear example and holds quite generally: there exists a lending contract $(b_t,R_{t+1})$ that can replicate the payoff from selling the asset: just set $b_t=\E[p_{t+1}|p_t]$ and make $R_{t+1}$ infinitely large, so that at $t+1$ the agent defaults under all circumstances. Due to the martingale property of the price process, we have $\E[p_{t+1}|p_t]=p_t$, so that the sum which the agent can raise through borrowing can be made as large as the sum from selling the asset.

At date 0, depending on the realization of $p_0$, the borrower would either prefer to borrow the sum $\hat{b}_1=2x$, which will be repaid with probability one at date $1$, or else, if $p_0$ is large enough, to borrow the sum
\begin{equation*}
\hat{b}_0=2p_0+x+\frac{x^2}{8p_0(1-\beta)},
\end{equation*}
an amount on which he will default for sure at $t=1$. Such a risky loan contract is, in fact, equivalent to selling the asset, which is the interpretation given in the proposition.

Next, it should be noted that the bang-bang nature of the solution is due to linearity of the utility function. For concave utility, one would anticipate that in each period, the range of asset prices can be partitioned into the three regions: the agent would engage in risky borrowing for very large prices, would keep the asset and borrow safely (that is, borrow the amount that will be repaid on the next date under all circumstances) for the intermediate range of asset prices, and would default on the existing debt if the asset price happens to fall below the specified threshold.

To summarize the discussion, for the case when the individual is known to be strategic and can borrow against the asset, the agent sells the asset at date 0 if its price happens to be larger than the threshold; otherwise, he borrows safely and keeps the asset until date 1, in which case he borrows against the asset, and defaults at date 2 if the asset price drops below the threshold.

In this version of the model, defaults would be observed only at date 2, and only in case of substantial increase in the asset price from $p_0$ to $p_1$ and the subsequent decrease from $p_1$ to $p_2$, that is, when the asset price follows the hump-shaped pattern. To the extent that such a pattern in $\{p_t\}$ can be attributed to bubbles,\footnote{Since the price process is postulated exogenously, this setup remains silent about the possible cause of the rise in the asset price above the fundamental in the first place.} the model highlights one channel through which a burst of the bubble triggers default.

\subsection{Infinite horizon}

Let us analyze the setup in which the time horizon may be arbitrarily long, but is finite, and consists of $T+1$ dates.\footnote{This counter includes date 0, so that $T$ is the last date. For instance, in the canonical model developed in the previous section, we have $T=2$.} Take $\tau\in\{1,\ldots,T\}$ to be the last date following non-autarkic continuation game -- that is, the game with $(b_{\tau-1},R_{\tau})\neq(0,0)$.

In what follows, we consider the general process for the asset price, described by the function $F(\cdot|p)$, which satisfies the martingale property:
\begin{equation*}
\E[p_{t+1}|p_t]\equiv\int p_{t+1}d  F(p_{t+1}|p_t)=p_t
\end{equation*}

First, consider the case when the asset has already been sold by date $\tau$: $a_{\tau}=0$. The continuation value of the honest type is given by
\begin{equation*}
V_{\tau}(R_{\tau};\pi_{\tau},0)=\max_{b_{\tau},R_{\tau+1}}\Big\{y_{\tau}-R_{\tau}+b_{\tau}+\beta(y_{\tau+1}-R_{\tau+1}+V_{\tau+1})\Big\},
\end{equation*}
subject to the zero-profit condition:
\begin{equation*}
b_{\tau}=\pi_{\tau}R_{\tau+1}
\end{equation*}
and the feasibility constraint\footnote{That is, the non-negativity of date-$\tau$ consumption: $c_{\tau}=y_{\tau}-R_{\tau}+b_{\tau}\geq0$.}
\begin{equation*}
b_{\tau}\geq R_{\tau}-y_{\tau}.
\end{equation*}

In order for the honest type to be willing to stop borrowing at date $\tau$, it must be the case that
\begin{equation*}
\pi_{\tau}\leq\beta.
\end{equation*}

At the same time, in order to be willing to offer non-autarkic contract at date $\tau-1$, it must be the case that
\begin{equation*}
\pi_{\tau-1}\geq\beta.
\end{equation*}

However, since the sequence of the lenders' posterior is non-decreasing, we must have
\begin{equation*}
\pi_{\tau-1}\leq\pi_{\tau},
\end{equation*}
because each period in which the borrower repays the outstanding debt, the lenders' posterior that the type is honest must increase or stay the same.

Clearly, the above three inequalities cannot hold simultaneously. Therefore, once the honest type decides to sell the asset, he must realize that the subsequent consumption path will be autarkic: from the date $\tau$ onward, we have
\begin{equation*}
V_{\tau}^a=\sum_{t=0}^{T-\tau}\beta^ty_t.
\end{equation*}

This is intuitive, since due to the martingale property of the price process $\{p_t\}$, together with zero dividend stream ($x=0$), the asset is valuable only to the extent it allows the individual to increase future borrowing.

Now let us turn to the case where the individual holds the asset at date $\tau$. The borrower will be indifferent between selling the asset and keeping it whenever
\begin{equation*}
(\pi_{\tau}-\beta)y_{\tau+1}+\beta\int\max\Big\{(\pi_{\tau+1}-\beta)y_{\tau+2},p_{\tau+1}\Big\}d  F(p_{\tau+1}|p_{\tau})=p_{\tau}.
\end{equation*}

The right-hand side of the above expression is the net gain from selling the asset in period $\tau$. The left-hand side is the net gain from borrowing against the asset in period $\tau$ (the term $y_{\tau+1}$ represents the maximal sum the honest borrower will be able to borrow, given the feasibility constraint), plus the option value of holding the asset until the next date.

The behavior of the strategic type pins down the evolution of the lenders' posterior beliefs $\{\pi_t\}_{t=\tau}^T$, jointly with the indifference condition of the strategic type (defaulting now versus repaying and defaulting on a greater debt in the future).

For each period $t$, the set of prices in period $\tau$ can be subdivided into two subsets: $\mathcal{P}_t^h$ and its complement, $\overline{\mathcal{P}}_t^h$, for which the honest type prefers to keep and, respectively, to sell the asset. In the specific case analyzed in the previous section, the set $\mathcal{P}_t^h$ was a closed interval at $t=0$, and was empty for $t=1$ and 2: the honest type \emph{always} preferred to sell the asset at the interim stage 1. For the general structure outlined here (that is, for the horizon length $T\geq2$ and for arbitrary distribution $F$), the set $\mathcal{P}_t^h$ may take a more complex structure, and may vary from one date to the other.

Similar reasoning can be applied to the two disjoint subsets $\mathcal{P}_t^s$ and $\overline{\mathcal{P}}_t^s$, for which the strategic type prefers to repay the previous-period debt versus default on it, respectively. Although it is true that, other things equal, the strategic type would prefer to repay on the loan and keep the asset as $p_t$ increases, because the option value of selling the asset rises -- at the same time, the option value of keeping the asset and defaulting on it in the future may decrease with the rise in $p_t$.

Therefore, although the length of the time horizon can be easily increased to length exceeding three periods, the assumption of the uniform distribution for the price process is not innocuous and cannot be dropped altogether: the interval construction of Figure \ref{eqstruc} is sensitive to the specifics of the uniform distribution.

\subsection{Overlapping generations}

In order to disentangle the dynamics driven by asset price changes from the individual's life-cycle considerations, we may consider the following generalized version of the setup: the time horizon will be infinite: $t=0,1,2,\ldots$, but at each date $t$, there will be three generations of borrowers: young ($0$), middle-aged ($1$) and old ($2$).

Old generation at period zero (the generation that was born at date $t=-2$) will be assumed to hold one unit of the asset, whereas the young and middle-aged individuals will be assumed to earn non-financial incomes $y^0$ and $y^1$, with $y^1\geq y^2$. Asset can be traded among the generations, and can also be liquidated by the lenders in case of default.

In addition, each newly born generation of borrowers will contain the share $\pi_0$ of the honest borrower types, as well as $1-\pi_0$ of the strategic types.

The date-$t$ realization of the asset price leads to different consequences for the borrowers at the different stages of their life cycle (that is, for different generations). The structure of equilibrium illustrated on Figure \ref{eqstruc} suggests that for the oldest generation, a sudden drop in the market price would lead to default of the entire sub-population of the mass $1-\pi_0$, the strategic types. If the \emph{level} of the asset price remains relatively high, however, only \emph{some} of the middle-aged borrowers will choose to default, whereas if the asset price is very low, their default will be deterministic.

At the same time, the entire cohort of the middle-aged borrowers who didn't default on their previous-period loan will choose to sell the asset and either remain in autarky if the current asset price is relatively high, or choose to borrow against their reputation -- in the opposite case. Interestingly, one would be able to observe \emph{credit rationing} for the middle-aged group in case if the current asset price lies in the interval $[R_1-\beta y_2,\overline{p}_1]$, that is, when it is quite high but not overly so.

As for the youngest generation, the model predicts massive selling of the asset when its price is either overly high or overly low, whereas moderate price level is required for the asset-backed borrowing to emerge in the first place. In particular, defaults by the middle-aged borrowers can co-exist with borrowing by the young cohort.

\subsection{Robustness and generalizations}

Since we have presented a very specific environment, in this section, we discuss to what extent our main results are dependent on the specific assumptions that we have made and what the dimensions are along which our setting can be generalized. We also discuss how our insights can be applied to a broader context outside the lender-borrower relationship.

Nothing important hinges on the assumption that $T=3$ -- the extension to a general $T$-period horizon would be straightforward.\footnote{The equilibrium characterized in Lemma \ref{alwayssell} always features asset sale at date $t=1$. But clearly, this is because period 2 is the last one. In the general $T$-period version, the borrower may keep the asset up to period $T-1$, at which point he prefers to sell because of complete separation at period $T$ (since the strategic type always defaults).} Uniform distribution of the prices allowed us to derive our results in closed form, but the main insight behind Proposition \ref{prop:borrow} does not rest upon it: as long as the conditional distribution for the price process satisfies the martingale property, the borrower will be inclined to sell the asset right away as long as the asset delivers zero non-pledgeable income ($x=0$). Likewise, I conjecture that the bang-bang feature of the solution characterized in Proposition \ref{prop:borrowasset} will generalize to a broader class of distributions.

On the other hand, the martingale property itself cannot be dispensed with that easily: as long as the borrower expects the asset's price to grow in the future, he may wish to postpone selling it, even if he does not intend to use it as collateral for borrowing. But to the extent that discounting is substantial ($\beta$ being well below one), the borrower's desire to shift consumption forward in time might outweigh his  desire to keep the asset and wait for a higher price.

The assumption that the asset price follows an exogenous stochastic process may seem harder to justify. We should stress that our paper's focus is on the \emph{partial equilibrium} analysis, where the borrower's behavior cannot have an impact on the determination of the equilibrium asset price.\footnote{\cite{Dubeyetal2005} have embedded default in the general-equilibrium context with incomplete markets. In the model of \cite{GeanakoplosZame2014}, equilibrium asset structure and price was endogenously determined.}

It would be insightful to see how the results of our model would differ if the asset price did not fluctuate. Suppose we had constant asset price:
\begin{equation*}
p_t=p\quad\text{for all $t$.}
\end{equation*}

Solving the model backwards, we first see that the individual's date-1 default and borrowing decision is independent of $p_0$. Hence, the properties of equilibrium constructed in the previous section will hold for the date-1 continuation game: specifically, if $p\in[0,R_1-y]$, the types will completely separate; there will be partial separation if we had $p\in[R_1-y,R_1]$, and there will be pooling for $p_1>R_1$. Likewise, at date 0, the borrower would prefer to keep the asset if we had $p\in[\underline{p}_0,\overline{p}_0]$.

This implies that the complementarity between collateral and reputation depends on asset pledgeability, and not on asset price fluctuations. However, these fluctuations are important: the analysis in section \ref{assetreputation} provides us some insights regarding comovement of asset prices and credit growth. In particular, it suggests that one would typically observe a wave of defaults when a rapid asset price increase is followed by a drastic decrease in asset prices. The analysis of the benchmark case of section \ref{onlyasset} highlighted the working of this mechanism when the asset was valuable for the stream of dividends $x>0$ for the borrowers. However, proposition \ref{prop:borrowasset} suggested that, if the individual does not have a chance to develop his reputation, it will either be the case that (i) he does not want to borrow in the first place (for $p_0\geq\hat{p}_0$, he prefers to sell the asset in period 0), or (ii) he will borrow, but will not default at date 1.

Introducing linear utility function with discounting was the easiest way to motivate the desire for borrowing. Perhaps a more conventional way would be to postulate the increasing and concave utility function $u(c_t)$, but the concavity would introduce the consumption-smoothing motive, and the optimal pattern of borrowing and lending would depend on the income stream  to a much greater extent. So, the main result of the paper would be much less clean in an environment with concave utility.

The assumption that the contract $(b_t,R_{t+1})$ is proposed by the borrower is harder to justify. An obvious alternative would be to assume the competitive screening setup, in which the lenders propose the contract to the borrower. If we suppose that the lenders' goal is to maximize the honest type's utility, then arguably the results from Section \ref{onlyreputation} (in particular, Lemma \ref{lemmasignal} and Proposition \ref{pardomcontr}) will be unchanged. The reason we decided not to introduce that assumption is the potential non-existence of equilibrium\footnote{This issue is the same as that which arises in the competitive screening setup, cf. \cite{RothschildStiglitz1976}.} when there is credit rationing -- in particular, this concerns the range for $p_1\in[R_1-\beta y_2,R_1]$ on Figure \ref{fig:lendpost}.

The assumption that the income stream $\{y_t\}_{t=1}^T$ is \emph{deterministic} allowed us to rule out the case when the borrower is \emph{forced} to default on a debt. Our choice to introduce this assumption is closely related to the way in which we modelled reputation for honesty: the honest type was assumed to be borrowers having \emph{infinite costs of default} irrespective for its reason. Clearly, this is not the only way to model honesty, and two alternative ways are worth mentioning.

First, one could assume that the borrower faces a privately observed cost of default $\theta>0$, and the lender only knows the distribution of $\theta$ but not its actual realization. In that case, the degree of the borrower's ``honesty'' may vary along the continuous scale with $\theta$. At each date, the borrower will default on the existing debt provided that his type $\theta$ is below a certain threshold $\theta^{*}(p_t,R_t)$, which depends on the current-period asset price and the outstanding debt. In that case, \emph{learning} about the borrower's type would occur in the truncation of the support of $\theta$: in every period when the borrower defaults, the lenders conclude that $\theta<\theta^{*}$. Such a framework would be an alternative model complementary to ours, and the extent to which our results would continue to hold in that kind of setup remains an open question.

Second, one may assume that the honest borrower finds it unacceptable to default \emph{strategically} but will not suffer any additional ``moral'' costs of default if he is \emph{forced} into default.\footnote{Technically, the default will be \emph{forced} whenever the current-period income $y_t$ plus the potential proceeds from the asset sale $p_t$ are insufficient to cover the outstanding debt $R_t$.} The interesting feature of this setup is that the lender sees the default but does not observe the borrower's current or past income realizations and hence is not sure whether the default is strategic or not -- and hence, the default does not destroy the borrower's reputation completely.\footnote{This possibility of redeeming oneself is broadly consistent with the evidence by \cite{Chatterjeeetal2011}, who show that the credit scores tend to be mean reverting.}

\section{\label{conclusion}Conclusion}

The paper's main focus is on the complementarity between external incentives to repay a loan (in the form of collateral), on the one hand, and internal incentives (in the form of reputation). 

It is shown that the possibility to pledge an asset as collateral in the loan contract makes it more attractive for the borrower to develop a reputation for being honest, that is: to borrow against the asset, repay the loan on the next date and, if the asset price did not increase substantially, to keep borrowing against increased reputation.

Theorem \ref{mainresult} shows the borrower's desire to develop a reputation for honesty -- in terms of the reduction of the lower bound for $\pi_0$ above which he prefers to keep the asset and borrow against it. The complementarity between secured and unsecured credit works whenever the ratio of financial to non-financial income, $\frac{p_0}{y}$, is neither too large nor too small.

In the concluding comments, we would like to discuss how the flavor of our results may relate to other contexts. Anecdotal evidence suggests that the popularity of opposition leaders in authoritarian regimes is largely affected by their willingness to make sacrifices for their principles. When the political climate deteriorates towards the abolition of civil liberties (an analogue of asset price drop in our setup), some political activists may give up their principles and fit into the regime (default), others may be forced into exile and continue their political struggle from abroad (selling), and yet others may stay in the country putting themselves in danger of imprisonment and even life threats (keeping the asset). Facing these threats creates the possibility of having their \emph{skin in the game} (to use the notion popularized by \cite{Taleb2018}), which enhances the working of the reputation mechanism.

\appendix

\section{\label{proofs}Proofs}

\subsection{\label{proof:borrow}Proof of Proposition \ref{prop:borrow}}

\begin{proof}
Suppose the agent has kept the asset and borrowed against it until date $2$. The gains to default will be that the borrower keeps $R_2$; the gains to repayment will be that he can sell the asset at a price $p_2$. Hence, at $t=2$, he will repay the debt provided that $p_2\geq R_2$.

In case the borrower holds the asset at the end of date 1, the contract $(b_1,R_2)$ that he will offer to the competitive lender will solve
\begin{equation}
\max_{b_1,R_2}\left\{-R_1+b_1+\frac{\beta}{2p_1}\int_{R_2}^{2p_1}(p_2-R_2)d  p_2\right\},
\end{equation}
subject to
\begin{equation}
b_1=\left(1-\frac{R_2}{2p_1}\right)R_2+\frac{1}{2p_1}\int_{0}^{R_2}p_2d  p_2.
\end{equation}

From the constraint equation, we get
\begin{equation*}
b_1=\left(1-\frac{R_2}{4p_1}\right)R_2,
\end{equation*}
which can be plugged into the maximization problem to yield:
\begin{equation*}
\hat{b}_1=p_1\quad\text{and}\quad \hat{R}_2=2p_1.
\end{equation*}

Evidently, the maximal possible realization of the date-$2$ asset price, $p_2$, is precisely equal to $p_2=2p_1$. Hence, promising the loan contract with $\hat{R}_2=2p_1$ means that at date $t=2$, there will be a default with probability one, since the default occurs whenever $p_2<\hat{R}_1=2p_1$, and with $p_2|p_1\sim\mathcal{U}[0,2p_1]$, we have $\Pr\{p_2<2p_1\}=1$. Hence, offering such a contract -- for the possibility to borrow the sum $\hat{b}_1=p_1$ is essentially \emph{equivalent} (in terms of the payoffs) to selling the asset at date $t=1$.

Now notice that the borrower's date-1 utility from repaying the previous-period debt ($R_1$) and borrowing against it is equal to
\begin{equation}
U_1^b(p_1,R_1)=-R_1+p_1,
\end{equation}
which is also equal to the utility from repaying the debt and selling the asset, $U_1^s(p_1,R_1)$.

On the other hand, if the borrower defaults, he will loose the asset and get zero utility:
\begin{equation}
U_1^d=0.
\end{equation}

Evidently, at date 1, the date-0 loan will be repaid provided that $U_1^b(p_1,R_1)\geq U_1^d$, which boils down to
\begin{equation*}
p_1\geq R_1.
\end{equation*}

Now consider the agent's borrowing decision at date 0.

Along the same lines, the contract $(b_0,R_1)$ solves
\begin{equation}
\max_{b_0,R_1}\left\{b_0+\frac{\beta}{2p_0}\int_{R_1}^{2p_0}U_1^b(p_1,R_1)d  p_1\right\},
\end{equation}
subject to
\begin{equation}
b_0=\left(1-\frac{R_1}{2p_0}\right)R_1+\frac{1}{2p_0}\int_{0}^{R_1}p_1d  p_1.
\end{equation}

Clearly, this is the same problem as the one for the date 1, except that all the time indices are shifted from $t$ to $t-1$ and we no longer have any outstanding debt $R_0$. If the agent were to borrow against the asset, the optimal contract that he will propose is
\begin{equation*}
\hat{b}_0=p_0\quad\text{and}\quad \hat{R}_1=2p_0,
\end{equation*}
which is equivalent to selling the asset.
\end{proof}

\subsection{\label{proof:lemmasignal} Proof of Lemma \ref{lemmasignal}}

\begin{proof}
First, consider the date-1 continuation game when there was no default, and denote the lenders' posterior that the borrower is of the `honest' type (conditional on no default) by $\pi_1\geq\pi_0$.

Any contract $(b_1,R_2)$ that satisfies three conditions,
\begin{equation*}
0\leq R_2\leq y_2,\quad b_1=\pi_1R_2\quad\text{and}\quad b_1+\beta(y_2-R_2)\geq\beta y_2,
\end{equation*}
is consistent with equilibrium.

The first inequality is the \emph{feasibility condition}: the borrower has to be able to repay the loan at date 2. The second is lenders' \emph{zero-profit condition}, and the third is the \emph{participation constraint} of the honest borrower. One contract that trivially satisfies these three conditions is, of course,
\begin{equation*}
b_1^{*}=R_2^{*}=0,
\end{equation*}
which is just non-participation.

The non-trivial contract exists only it $\pi_1$ happens to be large enough, namely if
\begin{equation*}
\pi_1\geq\beta.
\end{equation*}

If the above condition is satisfied, then any contract
\begin{equation}
\Big\{(b_1^{*},R_2^{*}):\quad b_1^{*}=\pi_1R_2^{*}\quad\text{and}\quad0\leq R_2^{*}\leq y_2\Big\}
\end{equation}
is consistent with equilibrium.

Now consider the strategic borrower's choice of whether to default at date 1.

He is \emph{indifferent} between defaulting and repaying, provided that
\begin{equation*}
y_1-R_1+b_1^{*}+\beta y_2=y_1+\beta y_2,
\end{equation*}
so that
\begin{equation*}
b_1^{*}=R_1.
\end{equation*}

That is to say, the borrower has to be able to exactly roll over his debt on to the next period. If his date-1 borrowing capacity is strictly larger than the repayment on the previous-period loan ($b_1^{*}>R_1$), he will never default ($\delta_1=0$). If on the other hand he has to repay more than what he can raise ($b_1^{*}<R_1$), he will default with certainty.

Therefore, for a given $b_1^{*}$, the probability of default as a function of $R_1$, is equal to
\begin{equation}
\delta_1(R_1;b_1^{*})
\begin{cases}
=0&\text{if }R_1<b_1^{*},\\
\in[0,1]&\text{if }R_1=b_1^{*},\\
=1&\text{if }R_1>b_1^{*}.
\end{cases}
\end{equation}

Consequently, the lenders' posterior probability that the borrower is of the honest type is given by
\begin{equation}
\pi_1(R_1;b_1^{*},\pi_0)
\begin{cases}
=\pi_0&\text{if }R_1<b_1^{*},\\
\in[\pi_0,1]&\text{if }R_1=b_1^{*},\\
=1&\text{if }R_1>b_1^{*}.
\end{cases}
\end{equation}

Now consider the borrowing decision at date 0. Given the strategic borrower's default probability $\delta_1(R_1;b_1^{*})$ at date 1, the date-0 lenders' zero-profit condition is given by
\begin{equation*}
b_0^{*}=\Big[\pi_0+\big(1-\pi_0\big)\big(1-\delta_1(R_1;b_1^{*})\big)\Big]R_1.
\end{equation*}

Given the Bayes' rule for the lenders' posterior updating,
\begin{equation}
\pi_1(R_1;b_1^{*},\pi_0)=\frac{\pi_0}{\pi_0+(1-\pi_0)(1-\delta_1(R_1;b_1^{*}))},
\end{equation}
the zero-profit condition can be rewritten as
\begin{equation}
b_0(R_1;b_1^{*},\pi_0)=\frac{\pi_0}{\pi_1(R_1;b_1^{*},\pi_0)}R_1.
\end{equation}

Now, the participation constraint of the honest type is given by
\begin{equation*}
b_0+\beta(b_1^{*}-R_1)-\beta^2R_2^{*}\geq0,
\end{equation*}
or
\begin{equation*}
b_0\geq\beta R_1-\beta(b_1^{*}-\beta R_2^{*}).
\end{equation*}

Jointly with the lender's date-0 zero-profit condition, this yields
\begin{equation}
\left(\beta-\frac{\pi_0}{\pi_1(R_1;b_1^{*},\pi_0)}\right)R_1\leq\beta(b_1^{*}-\beta R_2^{*}).
\end{equation}

Since $\beta<1$, the above condition is satisfied for all $R_1<b_1^{*}$.

For $R_1=b_1^{*}$, the above condition is satisfied if and only if
\begin{equation*}
\beta^2\leq\pi_0.
\end{equation*}

For $R_1>b_1^{*}$, the above condition boils down to
\begin{equation*}
(\beta-\pi_0)R_1\leq\beta(1-\beta)b_1^{*},
\end{equation*}
which always holds for $\beta\leq\pi_0$, while for $\beta>\pi_0$, holds whenever
\begin{equation*}
R_1\leq\frac{\beta(1-\beta)}{\beta-\pi_0}b_1^{*}.
\end{equation*}

Finally, in order for the honest borrower to be able to repay the loan at date 1, the repayment must satisfy
\begin{equation*}
R_1\leq y_1+b_1^{*}.
\end{equation*}

We can now characterize an equilibrium at date 0 for a given continuation contract $(b_1^{*},R_2^{*})$ and for different ranges of $\pi_0$ and $\beta$.

First, consider the case when $\pi_0<\beta^2$.

We claim that the only possible equilibrium is autarky:
\begin{equation}
b_t^{*}=R_{t+1}^{*}=0,\quad\text{for $t=0,1$}.
\end{equation}

To see this, first notice that for the date-0 honest lender's zero-profit condition to be satisfied, it must be the case that
\begin{equation*}
R_1^{*}<b_1^{*}.
\end{equation*}

However, we already know that the strategic type will never default on such a contract, and hence there will be no update of beliefs at date 1:
\begin{equation*}
\pi_1=\pi_0.
\end{equation*}

Next, since $\beta<1$, we have
\begin{equation*}
\pi_1=\pi_0<\beta^2<\beta,
\end{equation*}
and thus the only contract available at date 2 is
\begin{equation*}
b_1^{*}=R_2^{*}=0.
\end{equation*}

Jointly with the requirement that
\begin{equation*}
R_1^{*}\leq\frac{\beta(1-\beta)}{\beta-\pi_0}b_1^{*},
\end{equation*}
this yields
\begin{equation*}
b_0^{*}=R_1^{*}=0.
\end{equation*}

Now consider the case when $\beta^2\leq\pi_0<\beta$.

First, we claim that we cannot have an equilibrium in which
\begin{equation*}
R_1^{*}<b_1^{*}.
\end{equation*}

By contradiction, suppose this were the case. Then we would have $\pi_1=\pi_0<\beta$, implying that only autarkic equilibrium ($b_1^{*}=R_2^{*}=0$) is possible at date 1. Hence, we must have $R_1^{*}\geq b_1^{*}$. When the inequality is strict, the strategic borrower defaults with probability one at date 1, and hence
\begin{equation*}
\pi_1(R_1^{*};b_1^{*},\pi_0)=1\quad\text{and}\quad b_0^{*}=\pi_0R_1^{*}.
\end{equation*}
so that the date-1 debt, conditional on the previous repayment, is riskless:
\begin{equation*}
b_1^{*}=R_2^{*}.
\end{equation*}

So, any contract $(b_0^{*},R_1^{*})$ with
\begin{equation}
b_0^{*}=\pi_0R_1^{*}\leq\frac{\beta(1-\beta)}{\beta-\pi_0}\pi_0b_1^{*}\quad\text{and}\quad R_1^{*}\leq y_1+b_1^{*}
\end{equation}
is consistent with equilibrium.

In case of equality, $R_1^{*}=b_1^{*}$, we have a class of equilibria with
\begin{equation*}
b_0^{*}=\pi_0R_2^{*},
\end{equation*}
with the strategic borrower randomizing between repaying and defaulting, with
\begin{equation}
\delta_1^{*}(b_1^{*},R_2^{*};\pi_0)=1-\frac{\pi_0}{1-\pi_0}\left(\frac{R_2^{*}}{b_1^{*}}-1\right).
\end{equation}

Finally, take the case when $\pi_0\geq\beta$.

Now the only binding constraint is the date 1 is feasibility. In addition to the two equilibria analyzed for the case when $\pi_0\in[\beta^2,\beta)$, there exists another one in which the strategic borrower \emph{never} defaults at date 1: in that case, we must have
\begin{equation*}
b_1^{*}=\pi_0R_2^{*},
\end{equation*}
and any contract $(b_0^{*},R_1^{*})$ with
\begin{equation*}
b_0^{*}=R_1^{*}\leq y_1+b_1^{*}
\end{equation*}
is consistent with equilibrium.
\end{proof}

\subsection{\label{proof:pardomcontr}Proof of Proposition \ref{pardomcontr}}

\begin{proof}
Start with the date-1 continuation game after there was a repayment on date-0 loan.

Given $R_1$, the lenders' zero profit condition is
\begin{equation}
b_1(R_2;R_1,\pi_0)=
\begin{dcases}
R_2&\text{if }R_2<R_1,\\
R_1&\text{if }R_2\in\left[R_1,\tfrac{R_1}{\pi_0}\right].\\
\pi_0R_2&\text{if }R_2>\tfrac{R_1}{\pi_0}.
\end{dcases}
\end{equation}

This piecewise linear function describes the solid black line on Figure \ref{fig:feasdate}(a).

Depending on the value of $\beta$, there are two potential candidates for the honest borrower's optimum in the date-1 continuation game.

First, the borrower may propose the contract $(b_1,R_2)=(R_1,R_1)$. This is the contract with the risk-free interest rate, and the lenders will accept such a contract only if repayment of the date-0 loan induces their posterior update to $\pi_1=1$. In turn, this update will be consistent with equilibrium only if the strategic type defaults with probability one. The lending contract with $b_1=R_1$ allows the borrower who repays his date-0 debt to borrow exactly the same amount at date 1 -- and the strategic type will be \emph{indifferent} between repaying his date-0 debt $R_1$, borrowing $b_1=R_1$ and defaulting on this loan at date 2, on the one hand, and defaulting on his date-0 loan, on the other. Of course, any randomization between default and repayment will also be optimal. However, for the lenders' posterior to jump up to $\pi_1=1$, it must be the case that the strategic type defaults with probability one on his date-0 loan. This equilibrium involves a specific tie-breaking rule in the strategic borrower's default choice.

Alternatively, the honest type may prefer to choose the pooling contract $(b_1,R_2)=(\pi_0y_2,y_2)$ (in case if $b_1>R_1$, the strategic type will prefer to repay his date-0 loan and pool with the honest type).

At date $t=1$, the borrower is exactly indifferent between the two contracts whenever
\begin{equation}
R_1-\beta R_1=\pi_0y_2-\beta y_2.
\end{equation}

Whenever $\pi_0\geq\beta$, this equation has a non-negative solution
\begin{equation}
R_1=\frac{(\pi_0-\beta)y_2}{1-\beta}.
\end{equation}

Denote $\overline{R}_1=\max\left\{\frac{(\pi_0-\beta)y_2}{1-\beta},0\right\}$.

Now consider the situation at date 0. The date-0 lenders anticipate that at date 1, the borrower (having repaid the date-0 debt) will offer the contract $(b_1^{*},R_2^{*})$. Given the date-1 default decision of the strategic type, the date-0 lenders' zero-profit condition will be given by
\begin{equation*}
b_0(R_1;\overline{R}_1)
\begin{dcases}
=R_1&\text{if }R_1<\overline{R}_1,\\
\in\left[\pi_0\overline{R}_1,\overline{R}_1\right]&\text{if }R_1=\overline{R}_1,\\
=\pi_0R_1&\text{if }R_1>\overline{R}_1.
\end{dcases}
\end{equation*}

The two potential candidates for the honest borrower's optimum are: (i) propose the pooling contract $(b_0,R_1)=(\overline{R}_1,\overline{R}_1)$, which would allow to exactly roll-over the debt at date 1. In order to be willing to accept this contract at a risk-free rate, the date-0 lenders have to believe that the strategic borrower will repay the debt with probability one.

Alternatively,  the honest borrower may propose the separating contract
\begin{equation}
(b_0,R_1)=(\pi_0(\overline{R}_1+y_1),\overline{R}_1+y_1),
\end{equation}
which would trigger default by the strategic type at date 1.

The honest borrower's date-0 utility from the pooling contract is
\begin{equation}
U_{\text{pooling}}=(1-\beta^2)\overline{R}_1,
\end{equation}
whereas the utility from the separating contract is
\begin{equation}
U_{\text{separating}}=(\pi_0-\beta)y_1+(\pi_0-\beta^2)\overline{R}_1
\end{equation}

The separating contract (the one that separates the types early, at date $t=1$) will be preferable whenever $U_{\text{separating}}>U_{\text{pooling}}$, which (after invoking the expression for $\overline{R}_1$) boils down to
\begin{equation}
\pi_0>1-(1-\beta)\frac{y_1}{y_2}.
\end{equation}

Defining $g=\frac{y_2-y_1}{y_1}$, we have $\frac{y_1}{y_2}=\frac{1}{1+g}$, which yields the result.
\end{proof}

\subsection{\label{proof:alwayssell}Proof of Lemma \ref{alwayssell}}

\begin{proof}
Denote the lenders' posterior by
\begin{equation*}
\pi_1=\pi_1(p_1;R_1).
\end{equation*}

If the honest borrower decides to keep the asset and borrow against it, he would choose $(b_1,R_2)$, solving
\begin{equation}
\max_{b_1,R_2}\bigg\{(1+\beta)y-R_1+b_1+\beta(p_1-R_2)\bigg\},
\end{equation}
subject to the lenders' zero-profit condition:
\begin{equation}
b_1=
\begin{dcases}
R_2-(1-\pi_1)\frac{R_2^2}{4p_1},&\text{if }R_2<2p_1,\\
\pi_1R_2+(1-\pi_1)p_1,&\text{if }R_2\geq2p_1.
\end{dcases}
\end{equation}

As can be easily inspected, the optimal contract proposed by the honest type is given by
\begin{equation}
(b_1^{*},R_2^{*})=
\begin{dcases}
\left(\frac{1-\beta^2}{1-\pi_1}p_1,\frac{2(1-\beta)}{1-\pi_1}p_1\right)&\text{if }\pi_1<\beta,\\
\bigg(\pi_1y+(1-\pi_1)p_1,y\bigg)&\text{if }\pi_1\geq\beta.
\end{dcases}
\end{equation}

The honest type's utility from borrowing against the asset it thus given by
\begin{equation}
U^h_b(p_1;R_1,\pi_1)=(1+\beta)y-R_1+\beta p_1+
\begin{dcases}
\frac{(1-\beta)^2}{1-\pi_1}p_1,&\text{if }\pi_1<\beta,\\
(\pi_1-\beta)y+(1-\pi_1)p_1,&\text{if }\pi_1\geq\beta.
\end{dcases}
\end{equation}

If instead the honest type were to sell the asset and borrow against reputation, his optimal contract would be
\begin{equation}
(b_1^{*},R_2^{*})=
\begin{dcases}
(0,0)&\text{if }\pi_1<\beta,\\
(\pi_1y,y)&\text{if }\pi_1\geq\beta.
\end{dcases}
\end{equation}

Correspondingly, the honest type's utility from selling the asset is given by
\begin{equation}
U^h_s(p_1;R_1,\pi_1)=(1+\beta)y-R_1+p_1+
\begin{dcases}
0,&\text{if }\pi_1<\beta,\\
(\pi_1-\beta)y,&\text{if }\pi_1\geq\beta.
\end{dcases}
\end{equation}

Comparing the utility from selling the asset to the utility from keeping and borrowing against it on these two segments separately, we see that \emph{selling the asset at date 1 and borrowing exclusively against reputation} dominates keeping the asset for the honest type.

Given $R_1$ and $\pi_1$, the honest type's continuation utility is equal to
\begin{equation}
U^h_1(p_1)=(1+\beta)y-R_1+p_1+\max\Big\{(\pi_1-\beta)y,0\Big\}.
\end{equation}
\end{proof}

\subsection{\label{proof:initdef}Proof of Proposition \ref{initdef}}

\begin{proof}
If the strategic type defaults on the loan, he loses both the asset and reputation, thereby consuming only his non-financial income, and his autarkic continuation utility is given by $(1+\beta)y$.

If he instead mimics the honest type, repays the debt and sells the asset, his continuation utility is given by
\begin{equation*}
U^s_b(p_1;R_1,\pi_1)=(1+\beta)y-R_1+p_1+
\begin{dcases}
0&\text{if }\pi_1<\beta,\\
\pi_1y&\text{if }\pi_1\geq\beta.
\end{dcases}
\end{equation*}

As can be seen, whenever $p_1>R_1$, the strategic type repays the debt irrespective of $\pi_1$, and thus we have
\begin{equation*}
\pi_1=\pi_0<\beta.
\end{equation*}

Next, he is \emph{indifferent} between defaulting and repaying whenever
\begin{equation*}
\pi_1=\frac{R_1-p_1}{y}\quad\text{and}\quad\pi_1\geq\beta.
\end{equation*}

Overall, the strategic type's randomization leads to the following lenders' posterior by the end of date 1:
\begin{equation}
\pi_1(p_1;R_1)=
\begin{dcases}
1&\text{if }p_1<R_1-y,\\
\frac{R_1-p_1}{y}&\text{if }R_1-y\leq p_1<R_1-\beta y,\\
\beta&\text{if }R_1-\beta y\leq p_1\leq R_1,\\
\pi_0&\text{if }p_1>R_1.
\end{dcases}
\end{equation}

It should be noted that, in order for the posterior to be kept at $\beta$ for the price range
\begin{equation*}
\pi_1\in[R_1-\beta y,R_1],
\end{equation*}
the \emph{lenders have to be randomizing} as well: namely, whenever the borrower proposes the contract $(\pi_1y,y)$, the lenders should accept this contract with probability $\alpha(p_1;R_1)$, so as to make the strategic borrower \emph{indifferent} between defaulting and repaying:
\begin{equation}
\alpha(p_1;R_1)=\frac{R_1-p_1}{\beta y}.
\end{equation}

Given Bayes' rule for the update on behalf of lenders:
\begin{equation}
\pi_1(p_1;R_1)=\frac{\pi_0}{\pi_0+(1-\pi_0)(1-\delta_1(p_1;R_1))},
\end{equation}
we can back out the strategic borrower's equilibrium default probability:
\begin{equation}
\delta_1(p_1;R_1)=
\begin{dcases}
1&\text{if }p_1<R_1-y,\\
1-\frac{\pi_0}{1-\pi_0}\left(\frac{y_2}{R_1-p_1}-1\right)&\text{if }R_1-y\leq p_1<R_1-\beta y,\\
\frac{\beta-\pi_0}{\beta(1-\pi_0)}&\text{if }R_1-\beta y\leq p_1\leq R_1,\\
0&\text{if }p_1>R_1.
\end{dcases}
\end{equation}

The price thresholds are thus given by
\begin{equation}
\underline{p}_1=R_1-y\quad\text{and}\quad\overline{p}_1=R_1.
\end{equation}

Correspondingly, the honest type's date-1 utility is given by
\begin{equation}
U^h_1(p_1;R_1)=(1+\beta)y-R_1+p_1+
\begin{dcases}
(1-\beta)y,&\text{if }p_1<R_1-y,\\
R_1-p_1-\beta y,&\text{if }R_1-y\leq p_1\leq R_1-\beta y,\\
0,&\text{if }p_1>R_1-\beta y,
\end{dcases}
\end{equation}
which can be rewritten as
\begin{equation}
U^h_1(p_1;R_1)=
\begin{dcases}
2y-R_1+p_1,&\text{if }p_1<R_1-y,\\
y,&\text{if }R_1-y\leq p_1\leq R_1-\beta y,\\
(1+\beta)y-R_1+p_1,&\text{if }p_1>R_1-\beta y.
\end{dcases}
\end{equation}

\end{proof}

\subsection{\label{proof:mainresult}Proof of Theorem \ref{mainresult}}

\begin{proof}
Since we consider the case when $\pi_0<\beta$, if the borrower were to sell the asset, there will be autarkic equilibrium in the corresponding continuation game, with the associated utility of
\begin{equation}
\label{eq:utsell}
U_1^s=\beta(1+\beta)y+p_0.
\end{equation}

If the honest borrower were to keep the asset, he would optimally choose the contract $(b_0,R_1)$, solving
\begin{equation}
\label{combor0}
\max_{b_0,R_1}\left\{b_0+\frac{\beta}{2p_0}\int_0^{2p_0}U^h_1(p_1;R_1)d  p_1\right\},
\end{equation}
subject to the lenders' zero-profit condition:
\begin{equation}
\label{zeropr0}
b_0=\frac{1}{2p_0}\int_0^{2p_0}\left\{\frac{\pi_0}{\pi_1(p_1;R_1)}R_1+\left(1-\frac{\pi_0}{\pi_1(p_1;R_1)}\right)p_1\right\}d  p_1
\end{equation}
and the feasibility constraint telling that at date 2, the borrower should be \emph{able} to repay the loan under all circumstances:
\begin{equation}
\label{feasib0}
R_1\leq y+p_1+b_1^{*}(p_1;R_1),\quad\text{for all $p_1$}.
\end{equation}

This constraint has to hold for all $p_1$, and it can be shown that the worst-case scenario is the lowest realization $p_1=0$, in which case the constraint translates into
\begin{equation*}
R_1\leq2y.
\end{equation*}

We can write the honest borrower's objective function as
\begin{equation}
U_0^h(R_1;p_0)=\frac{1}{2p_0}\int_0^{2p_0}\underbrace{\left\{\frac{\pi_0}{\pi_1(p_1;R_1)}R_1+\left(1-\frac{\pi_0}{\pi_1(p_1;R_1)}\right)p_1+\beta U_1^h(p_1;R_1)\right\}}_{v_0^h(p_1;R_1)}d  p_1,
\end{equation}
where the integrand is given by
\begin{equation}
v_0^h(p_1;R_1)=
\begin{dcases}
2\beta y-(\beta-\pi_0)R_1+(1+\beta-\pi_0)p_1&\text{if }0\leq p_1<R_1-y,\\
(\beta+\pi_0)y+p_1&\text{if }R_1-y\leq p_1<R_1-\beta y,\\
\beta(1+\beta)y+\left(\frac{\pi_0}{\beta}-\beta\right)R_1+\left(1+\beta-\frac{\pi_0}{\beta}\right)p_1&\text{if }R_1-\beta y\leq p_1<R_1,\\
\beta(1+\beta)y+(1-\beta)R_1+\beta p_1&\text{if }R_1\leq p_1\leq2p_0.
\end{dcases}
\end{equation}

Integrating this function over $p_1$ and simplifying, we eventually obtain
\begin{equation}
\begin{split}
U_0^h(R_1;p_0)=&-\frac{(1-\pi_0)}{4p_0}R_1^2+(1-\beta)\left(1+\frac{\beta y}{2p_0}\right)R_1\\
&+\frac{(1-\beta)(\pi_0-\beta(1+\beta))}{4p_0}y^2+\beta(1+\beta)y+\beta p_0
\end{split}
\end{equation}

This function has to be maximized subject to the constraint $R_1\leq2y$, hence the value of $R_1$ that maximizes this function is given by
\begin{equation}
R_1^{*}=\min\left\{\frac{(1-\beta)(2p_0+\beta y)}{1-\pi_0},2y\right\}.
\end{equation}

The constraint is binding whenever
\begin{equation}
\label{eq:condbindingmain}
y\leq\underbrace{\frac{p_0}{\frac{1-\pi_0}{1-\beta}-\frac{\beta}{2}}}_{\overline{y}}
\end{equation}
and is slack otherwise.

The maximized value of the borrower's objective is thus given by
\begin{equation*}
U_0^h(R_1^{*};p_0)=
\begin{dcases}
\tfrac{[\beta(3+\beta^2)+(5-\beta)\pi_0-4(1+\beta^2)]}{4p_0}y^2+(\beta^2-\beta+2)y+\beta p_0&\text{if }y\leq\overline{y},\\
\left(\tfrac{\beta^2(1-\beta)}{1-\pi_0}+\pi_0-\beta(1+\beta)\right)\tfrac{(1-\beta)y^2}{4p_0}+\left(\tfrac{\beta(1-\beta)^2}{1-\pi_0}+\beta(1+\beta)\right)y+\left(\beta+\tfrac{(1-\beta)^2}{1-\pi_0}\right)p_0&\text{if }y>\overline{y}.
\end{dcases}
\end{equation*}

A direct comparison of $U_0^h(R_1^{*};p_0)$ with $U_1^s$ reveals that keeping the asset at $t=0$ and borrowing against it would be preferable to selling the asset whenever the constraint $R_1\leq2y$ is binding, so that \eqref{eq:condbindingmain} is satisfied:
\begin{equation}
\label{bound1}
\pi_0\geq1-(1-\beta)\left(\frac{p_0}{y}+\frac{\beta}{2}\right),
\end{equation}
and at the same time the value of $U_0^h(R_1;p_0)$ evaluated at $R_1=2y$ is larger than $U_1^s$ as given by \eqref{eq:utsell}, which is satisfied for
\begin{equation}
\label{bound2}
\pi_0\geq\frac{4(1-\beta)}{(5-\beta)}\left(\frac{p_0}{y}-2\right)\frac{p_0}{y}+\frac{\beta(3+\beta^2)-4(1+\beta^2)}{5-\beta}.
\end{equation}

Taken together, these two inequalities yield the lower bound $\pi_0^{*}(\beta,p_0,y)$ as given by \eqref{eq:suffcond}. A direct comparison of \eqref{bound1} and \eqref{bound2} with $\beta$ reveals that this bound is less tight than $\beta$ whenever
\begin{equation}
2-\beta<\frac{2p_0}{y}<2+\sqrt{\left(\frac{8}{1-\beta}-\beta(2-\beta)\right)},
\end{equation}
which yields condition \eqref{eq:bounds}.

\end{proof}

\subsection{\label{proof:borrowasset}Proof of Proposition \ref{prop:borrowasset}}

\begin{proof}
At date 2, repayment will be made only if the gains from holding the asset, equal to the asset price plus the dividend, outweigh the gains to default, which equal $R_2$: that is, if the date-2 asset price exceeds the threshold:
\begin{equation*}
p_2\geq\underbrace{R_2-x}_{\hat{p}_2(R_2)}.
\end{equation*}

In case the borrower holds the asset at the end of date 1, he will offer the contract $(b_1,R_2)$, which solves
\begin{equation}
\max_{b_1,R_2}\left\{x-R_1+b_1+\frac{\beta}{2p_1}\int_{R_2-x}^{2p_1}(x+p_2-R_2)d  p_2\right\},
\end{equation}
subject to
\begin{equation}
b_1=\left(1-\frac{R_2-x}{2p_1}\right)R_2+\frac{1}{2p_1}\int_{0}^{R_2-x}p_2d  p_2.
\end{equation}

Expressing
\begin{equation*}
b_1=\left(1-\frac{R_2}{4p_1}\right)R_2+\frac{x^2}{4p_1}
\end{equation*}
from the constraint equation and maximizing with respect to $R_2$, we eventually get
\begin{equation}
\hat{b}_1=p_1+\frac{1-2\beta}{p_1}\left(\frac{x}{2(1-\beta)}\right)^2\quad\text{and}\quad \hat{R}_2=2p_1-\frac{\beta x}{1-\beta}.
\end{equation}

Observe that the loan is risky ($\hat{R}_2>x$) only if $p_1$ is sufficiently large, namely if
\begin{equation}
p_1\geq\frac{x}{2(1-\beta)},
\end{equation}
while in the opposite case, the individual would just prefer to borrow
\begin{equation}
\hat{b}_1=\hat{R}_2=x,
\end{equation}
repay the loan with certainty and realize his option to sell the asset at date 2.

So, the borrower's date-1 utility from repaying and holding the asset -- to borrow against it -- is equal to
\begin{equation}
U_1^b(p_1,R_1)=
\begin{dcases}
x-R_1+p_1+\frac{x^2}{4p_1(1-\beta)}&\text{if }p_1\geq\frac{x}{2(1-\beta)},\\
2x-R_1+\beta p_1&\text{if }p_1<\frac{x}{2(1-\beta)}.
\end{dcases}
\end{equation}

On the other hand, borrower's utility from repaying and selling the asset is equal to
\begin{equation}
U_1^s(p_1,R_1)=x-R_1+p_1,
\end{equation}
so that borrowing against the asset strictly dominates selling it.

Finally, the agent's utility to default is zero:
\begin{equation}
U_1^d=0,
\end{equation}
since he does not repay the loan issued at date 0, he loses the asset and can no longer borrow against it.

Therefore, at date 1, the loan will be repaid, if and only if
\begin{equation}
U_1^b(p_1,R_1)\geq0,
\end{equation}
that is, if and only if $p_1$ exceeds the threshold: $p_1\geq\hat{p}_1(R_1)$, where
\begin{equation}
\hat{p}_1(R_1)=
\begin{dcases}
0&\text{if }R_1\leq2x,\\
\frac{R_1-2x}{\beta}&\text{if }2x<R_1\leq2x+\frac{\beta x}{2(1-\beta)},\\
\frac{1}{2}\left[(R_1-x)+\sqrt{(R_1-x)^2-\frac{x^2}{1-\beta}}\right]&\text{if }R_1>2x+\frac{\beta x}{2(1-\beta)}.
\end{dcases}
\end{equation}

Now consider the agent's borrowing decision at date 0.

Along the same lines, the contract $(b_0,R_1)$ solves
\begin{equation}
\max_{b_0,R_1}\left\{b_0+\frac{\beta}{2p_0}\int_{\hat{p}_1(R_1)}^{2p_0}U_1^b(p_1,R_1)d  p_1\right\},
\end{equation}
subject to
\begin{equation}
b_0=\left[1-\frac{\hat{p}_1(R_1)}{2p_0}\right]R_1+\frac{1}{2p_0}\int_{0}^{\hat{p}_1(R_1)}p_1d  p_1.
\end{equation}

Let us rewrite the above problem in terms of the choice of the threshold $\hat{p}_1$:
\begin{equation*}
\max_{\hat{p}_1\in[0,2p_0]}\left\{\left(1-\frac{\hat{p}_1}{2p_0}\right)R_1(\hat{p}_1)+\frac{1}{2p_0}\int_{0}^{\hat{p}_1}p_1d  p_1+\frac{\beta}{2p_0}\int_{\hat{p}_1}^{2p_0}U_1^b(p_1,R_1(\hat{p}_1))d  p_1\right\},
\end{equation*}
where
\begin{equation*}
R_1(\hat{p}_1)=
\begin{dcases}
2x+\beta\hat{p}_1&\text{if }0<\hat{p}_1\leq\frac{x}{2(1-\beta)},\\
x+\hat{p}_1+\frac{x^2}{4\hat{p}_1(1-\beta)},&\text{if }\hat{p}_1\geq\frac{x}{2(1-\beta)}.
\end{dcases}
\end{equation*}

We perform the maximization with respect to $\hat{p}_1$ on two intervals separately.

First, take the interval
\begin{equation*}
\hat{p}_1\in\left[0,\frac{x}{2(1-\beta)}\right].
\end{equation*}

The first order condition yields:
\begin{equation*}
-\frac{2x+\beta\hat{p}_1}{2p_0}+\beta\left(1-\frac{\hat{p}_1}{2p_0}\right)+\frac{\hat{p}_1}{2p_0}-\frac{\beta^2}{2p_0}(2p_0-\hat{p}_1)=0,
\end{equation*}
which is a linear increasing function of $\hat{p}_1$.

Now take the interval where
\begin{equation*}
\hat{p}_1\in\left[\frac{x}{2(1-\beta)},2p_0\right].
\end{equation*}

The first order condition yields:
\begin{equation*}
\left(2p_0-\hat{p}_1\right)\left(1-\beta-\frac{x^2}{4\overline{p}_1^2}\right)+\hat{p}_1=0,
\end{equation*}
which is also monotonically increasing in $\hat{p}_1$ and positive for all $\hat{p}_1$ within the range.

Therefore, the solution is bang-bang: we have either
\begin{equation}
\hat{p}_1^{*}=0\quad\text{or}\quad\hat{p}_1^{*}=2p_0.
\end{equation}

To determine the optimal $\hat{p}_1^{*}$, we compute the value of the objective function at these two candidate solutions.

When $\hat{p}_1^{*}=2p_0$, we have
\begin{equation*}
U_0^b(p_0)=p_0,
\end{equation*}
whereas for $\hat{p}_1^{*}=0$, we have
\begin{equation*}
U_0^b(p_0)=2x+\beta^2p_0.
\end{equation*}

So, the optimal contract is given by
\begin{equation}
(\hat{b}_0,\hat{R}_1)=
\begin{cases}
(2x,2x)&\text{if }p_0<\hat{p}_0,\\
\left(p_0,2p_0+x+\frac{x^2}{8p_0(1-\beta)}\right)&\text{if }p_0\geq\hat{p}_0,
\end{cases}
\end{equation}
where
\begin{equation}
\hat{p}_0=\frac{2x}{1-\beta^2}.
\end{equation}
\end{proof}

\bibliographystyle{abbrvnat}
\bibliography{defaults}

\begin{thebibliography}{22}
\providecommand{\natexlab}[1]{#1}
\providecommand{\url}[1]{\texttt{#1}}
\expandafter\ifx\csname urlstyle\endcsname\relax
  \providecommand{\doi}[1]{doi: #1}\else
  \providecommand{\doi}{doi: \begingroup \urlstyle{rm}\Url}\fi

\bibitem[Chari et~al.(2014)Chari, Shourideh, and Zetlin-Jones]{Charietal2014}
V.~V. Chari, A.~Shourideh, and A.~Zetlin-Jones.
\newblock Reputation and persistence of adverse selection in secondary loan
  markets.
\newblock \emph{American Economic Review}, 104\penalty0 (12):\penalty0 4027 --
  4070, 2014.
\newblock ISSN 00028282.
\newblock URL
  \url{http://libproxy.mit.edu/login?url=http://search.ebscohost.com/login.aspx?direct=true&db=bth&AN=99777896&site=eds-live}.

\bibitem[Chatterjee et~al.(2011)Chatterjee, Corbae, and
  R{\i}os-Rull]{Chatterjeeetal2011}
S.~Chatterjee, D.~Corbae, and J.-V. R{\i}os-Rull.
\newblock A theory of credit scoring and competitive pricing of default risk.
\newblock \emph{Unpublished paper, University of Minnesota.[672]}, 31, 2011.

\bibitem[Cho and Kreps(1987)]{ChoKreps1987}
I.-K. Cho and D.~M. Kreps.
\newblock Signaling games and stable equilibria.
\newblock \emph{The Quarterly Journal of Economics}, 102\penalty0 (2):\penalty0
  179, 1987.
\newblock \doi{10.2307/1885060}.
\newblock URL \url{+ http://dx.doi.org/10.2307/1885060}.

\bibitem[Dell'Ariccia et~al.(2012)Dell'Ariccia, Igan, and
  Laeven]{DellAricciaetal2012}
G.~Dell'Ariccia, D.~Igan, and L.~Laeven.
\newblock Credit booms and lending standards: Evidence from the subprime
  mortgage market.
\newblock \emph{Journal of Money, Credit and Banking}, 44\penalty0
  (2-3):\penalty0 367--384, 2012.
\newblock ISSN 1538-4616.
\newblock \doi{10.1111/j.1538-4616.2011.00491.x}.
\newblock URL \url{http://dx.doi.org/10.1111/j.1538-4616.2011.00491.x}.

\bibitem[Dubey et~al.(2005)Dubey, Geanakoplos, and Shubik]{Dubeyetal2005}
P.~Dubey, J.~Geanakoplos, and M.~Shubik.
\newblock Default and punishment in general equilibrium1.
\newblock \emph{Econometrica}, 73\penalty0 (1):\penalty0 1--37, 2005.
\newblock \doi{https://doi.org/10.1111/j.1468-0262.2005.00563.x}.
\newblock URL
  \url{https://onlinelibrary.wiley.com/doi/abs/10.1111/j.1468-0262.2005.00563.x}.

\bibitem[Duffie et~al.(2002)Duffie, G{\^a}rleanu, and Pedersen]{Duffieetal2002}
D.~Duffie, N.~G{\^a}rleanu, and L.~H. Pedersen.
\newblock Securities lending, shorting, and pricing.
\newblock \emph{Journal of Financial Economics}, 66\penalty0 (2):\penalty0 307
  -- 339, 2002.
\newblock \doi{http://dx.doi.org/10.1016/S0304-405X(02)00226-X}.
\newblock Limits on Arbitrage.

\bibitem[Duffie et~al.(2007)Duffie, G{\^a}rleanu, and Pedersen]{Duffieetal2007}
D.~Duffie, N.~G{\^a}rleanu, and L.~H. Pedersen.
\newblock Valuation in over-the-counter markets.
\newblock \emph{The Review of Financial Studies}, 20\penalty0 (6):\penalty0
  1865--1900, 2007.

\bibitem[Edmans(2010)]{Edmans2010}
A.~Edmans.
\newblock The responsible homeowner reward: an incentive-based solution to
  strategic mortgage default.
\newblock 2010.

\bibitem[Elul et~al.(2010)Elul, Souleles, Chomsisengphet, Glennon, and
  Hunt]{Eluletal2010}
R.~Elul, N.~S. Souleles, S.~Chomsisengphet, D.~Glennon, and R.~Hunt.
\newblock What "triggers" mortgage default?
\newblock \emph{American Economic Review}, 100\penalty0 (2):\penalty0 490--94,
  May 2010.
\newblock \doi{10.1257/aer.100.2.490}.
\newblock URL \url{http://www.aeaweb.org/articles?id=10.1257/aer.100.2.490}.

\bibitem[Fay et~al.(2002)Fay, Hurst, and White]{Fayetal2002}
S.~Fay, E.~Hurst, and M.~J. White.
\newblock The household bankruptcy decision.
\newblock \emph{American Economic Review}, 92\penalty0 (3):\penalty0 706--718,
  June 2002.
\newblock \doi{10.1257/00028280260136327}.
\newblock URL
  \url{http://www.aeaweb.org/articles?id=10.1257/00028280260136327}.

\bibitem[Garmaise and Natividad(forthcoming)]{GarmaiseNatividad2017}
M.~J. Garmaise and G.~Natividad.
\newblock Consumer default, credit reporting, and borrowing constraints.
\newblock \emph{The Journal of Finance}, forthcoming.
\newblock ISSN 1540-6261.
\newblock \doi{10.1111/jofi.12522}.
\newblock URL \url{http://dx.doi.org/10.1111/jofi.12522}.

\bibitem[Geanakoplos and Zame(2014)]{GeanakoplosZame2014}
J.~Geanakoplos and W.~R. Zame.
\newblock Collateral equilibrium, i: A basic framework.
\newblock \emph{Economic Theory}, 56\penalty0 (3):\penalty0 443--492, 2014.
\newblock \doi{10.1007/s00199-013-0797-4}.
\newblock URL \url{https://doi.org/10.1007/s00199-013-0797-4}.

\bibitem[Guiso et~al.(2013)Guiso, Sapienza, and Zingales]{Guisoetal2013}
L.~Guiso, P.~Sapienza, and L.~Zingales.
\newblock The determinants of attitudes toward strategic default on mortgages.
\newblock \emph{The Journal of Finance}, 68\penalty0 (4):\penalty0 1473--1515,
  2013.
\newblock ISSN 1540-6261.
\newblock \doi{10.1111/jofi.12044}.
\newblock URL \url{http://dx.doi.org/10.1111/jofi.12044}.

\bibitem[Kreps and Wilson(1982)]{KrepsWilson1982}
D.~M. Kreps and R.~Wilson.
\newblock Reputation and imperfect information.
\newblock \emph{Journal of Economic Theory}, 27\penalty0 (2):\penalty0 253 --
  279, 1982.
\newblock ISSN 0022-0531.
\newblock \doi{http://dx.doi.org/10.1016/0022-0531(82)90030-8}.
\newblock URL
  \url{http://www.sciencedirect.com/science/article/pii/0022053182900308}.

\bibitem[Mailath et~al.(1993)Mailath, Okuno-Fujiwara, and
  Postlewaite]{Mailathetal1993}
G.~J. Mailath, M.~Okuno-Fujiwara, and A.~Postlewaite.
\newblock Belief-based refinements in signalling games.
\newblock \emph{Journal of Economic Theory}, 60\penalty0 (2):\penalty0
  241--276, 1993.
\newblock ISSN 0022-0531.
\newblock \doi{https://doi.org/10.1006/jeth.1993.1043}.
\newblock URL
  \url{https://www.sciencedirect.com/science/article/pii/S0022053183710434}.

\bibitem[Mian and Sufi(2009)]{MianSufi2009}
A.~Mian and A.~Sufi.
\newblock The consequences of mortgage credit expansion: Evidence from the u.s.
  mortgage default crisis*.
\newblock \emph{The Quarterly Journal of Economics}, 124\penalty0 (4):\penalty0
  1449, 2009.
\newblock \doi{10.1162/qjec.2009.124.4.1449}.
\newblock URL \url{+ http://dx.doi.org/10.1162/qjec.2009.124.4.1449}.

\bibitem[Mian and Sufi(2015)]{MianSufi2015}
A.~Mian and A.~Sufi.
\newblock \emph{House of debt: How they (and you) caused the Great Recession,
  and how we can prevent it from happening again}.
\newblock University of Chicago Press, 2015.

\bibitem[Milgrom and Roberts(1982)]{MilgromRoberts1982}
P.~Milgrom and J.~Roberts.
\newblock Predation, reputation, and entry deterrence.
\newblock \emph{Journal of Economic Theory}, 27\penalty0 (2):\penalty0 280 --
  312, 1982.
\newblock ISSN 0022-0531.
\newblock \doi{http://dx.doi.org/10.1016/0022-0531(82)90031-X}.
\newblock URL
  \url{http://www.sciencedirect.com/science/article/pii/002205318290031X}.

\bibitem[Ordo{\~n}ez et~al.(2019)Ordo{\~n}ez, Perez-Reyna, and
  Yogo]{Ordonezetal2019}
G.~Ordo{\~n}ez, D.~Perez-Reyna, and M.~Yogo.
\newblock Leverage dynamics and credit quality.
\newblock \emph{Journal of Economic Theory}, 2019.
\newblock ISSN 0022-0531.
\newblock \doi{https://doi.org/10.1016/j.jet.2019.06.001}.
\newblock URL
  \url{http://www.sciencedirect.com/science/article/pii/S0022053118301182}.

\bibitem[Rothschild and Stiglitz(1976)]{RothschildStiglitz1976}
M.~Rothschild and J.~Stiglitz.
\newblock Equilibrium in competitive insurance markets: An essay on the
  economics of imperfect information.
\newblock \emph{The Quarterly Journal of Economics}, 90\penalty0 (4):\penalty0
  629--649, 1976.

\bibitem[Taleb(2018)]{Taleb2018}
N.~Taleb.
\newblock \emph{Skin in the Game: Hidden Asymmetries in Daily Life}.
\newblock Incerto. Random House Publishing Group, 2018.
\newblock ISBN 9780425284636.
\newblock URL \url{https://books.google.ru/books?id=4dQ0DwAAQBAJ}.

\bibitem[Zingales(2010)]{Zingales2010}
L.~Zingales.
\newblock The menace of strategic default.
\newblock \emph{City Journal}, 20\penalty0 (2):\penalty0 47--51, 2010.

\end{thebibliography}

\end{document}